\documentclass[twocolumn]{article}
\usepackage[utf8]{inputenc}

\usepackage{amsmath}
\usepackage{amsfonts}
\usepackage{amssymb}
\usepackage{graphicx}
\usepackage{color}
\usepackage{multirow}
\usepackage{siunitx}

\usepackage[normalem]{ulem}

\begin{document}

\title{A topographic mechanism for arcing of dryland vegetation bands}
\author{Punit Gandhi\thanks{Mathematical Biosciences Institute, Ohio State University, Columbus, OH 43210. Email:gandhi.138@mbi.osu.edu},
Lucien Werner\thanks{Department of Computing and Mathematical Sciences, California Institute of Technology, Pasadena, CA 91125 }, 
Sarah Iams\thanks{John A. Paulson School of Engineering and Applied Sciences, Harvard University, Cambridge, MA 02138}, 
Karna Gowda\thanks{Department of Physics, University of Illinois at Urbana-Champaign, Urbana IL 61801}, 
Mary Silber\thanks{Committee on Computational and Applied Mathematics, and Department of Statistics, University of Chicago, Chicago IL 60637} }

\date{\today}

\maketitle

\begin{abstract} 
Banded patterns consisting of alternating bare soil and dense vegetation have been observed in water-limited ecosystems across the globe, often appearing along gently sloped terrain with the stripes aligned transverse to the elevation gradient. In many cases these vegetation bands are arced, with field observations suggesting a link between the orientation of arcing relative to the grade and the curvature of the underlying terrain. We modify the water transport in the Klausmeier model 
 of water-biomass interactions, originally posed on a uniform hillslope, to qualitatively capture the influence of terrain curvature on the vegetation patterns.  Numerical simulations of this modified model indicate that the vegetation bands change arcing-direction from convex-downslope when growing on top of a ridge to convex-upslope when growing in a valley. 
This behavior is consistent with observations from remote sensing data that we present here. Model simulations show further that whether bands grow on ridges, valleys, or both depends on the precipitation level. A survey of three banded vegetation sites, each with a different aridity level, indicates qualitatively similar behavior.

\vspace{5mm}
\noindent\textbf{Subject Areas:} biomathematics, biocomplexity, environmental science.
 
\end{abstract}

\section{Introduction}\label{sec:intro}

Self-organization of vegetation into community-scale spatial patterns has been observed in the drylands of five continents~\cite{borgogno2009mathematical, deblauwe2008global}. These distinctive patterns occur on a large spatial scale and may be monitored via satellite, so there has been growing interest in determining whether they hold information on the health of these ecosystems, or even provide an early warning sign of ecosystem collapse in response to desertification~\cite{rietkerk2004self,dakos2011slowing}. Given that only a handful of pattern characteristics can be reliably measured remotely, a more detailed understanding of the mechanisms that control these characteristics may help assess the potential for such ecosystem health predictions. 
Our modeling efforts, aimed at capturing the observed arcing-direction of vegetation bands relative to the direction of the mean elevation gradient, suggest the placement of vegetation patterns  on the terrain as a potential  indicator for aridity stress.  Higher precipitation level is required to support vegetation bands on the ridgelines than in the valleylines in the model, suggesting that vegetation bands confined to lower elevation channels may be the most vulnerable to collapse under increased aridity stress.  

\begin{figure*}
\includegraphics[width=\textwidth]{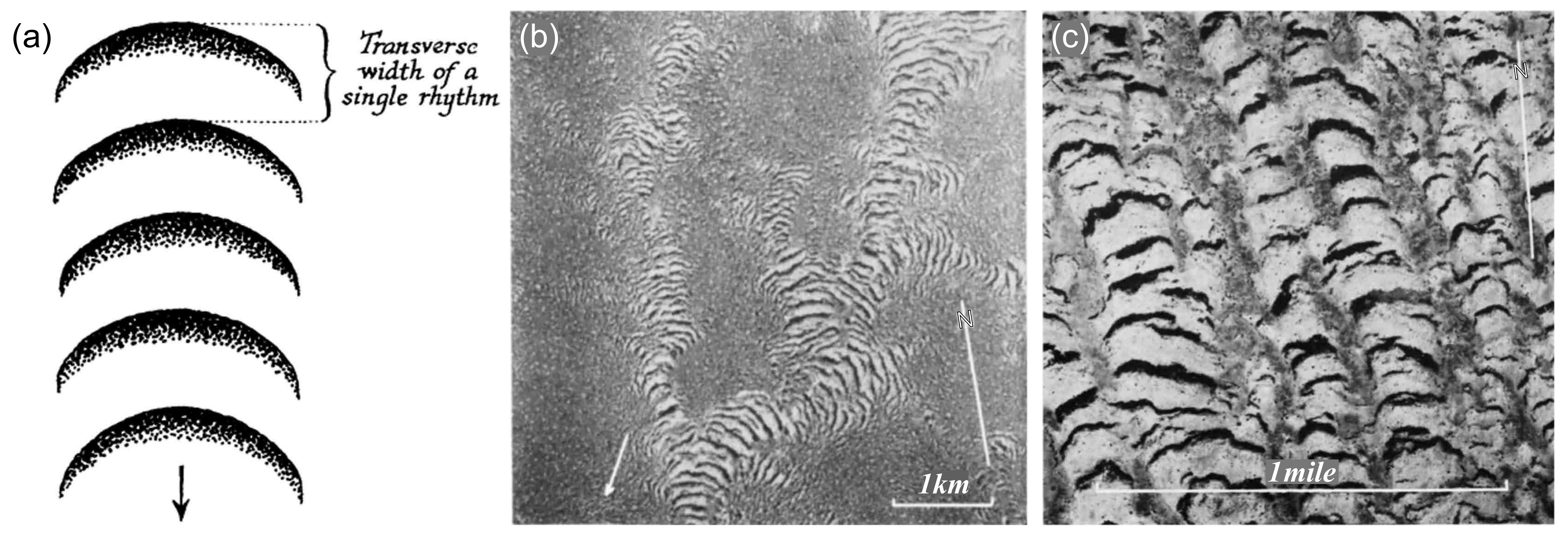}
\caption{   \label{fig:intro}  
Aerial photographs of vegetation patterns found in the Horn of Africa  alongside (a) a schematic taken from~\cite{macfadyen1950vegetation} showing typical orientation relative the topography. The vegetation bands are aligned transverse to the slope and arced convex-upslope.  The banded patterns in image (b), taken from~\cite{macfadyen1950vegetation}, appear within topographic depressions and the darker unpatterned regions of image (c), taken from~\cite{greenwood1957development}, are reported to be slightly elevated relative to the vegetation arcs. Arrows in (a) and (b) indicate downhill direction. }
\end{figure*}

Measured along the direction of the prevailing elevation gradient, a typical band pattern consists of patches of vegetation on the scale of tens of meters (i.e., band width), alternating with stretches of barren ground on the scale of tens to hundreds of meters, with this motif often repeating over a region of more than $10$ km in length (note scale bars in Fig.~\ref{fig:intro}).
Banded patterns undergo ecological dynamics, notably colonizing areas upslope to the bands during periods of plentiful rainfall, and retreating from downslope areas during periods of scarcity, resulting in a slow uphill migration of the bands over time~\cite{deblauwe2012determinants}. Measurements indicate that it would take almost a century for a vegetation band to travel its $\sim 30$ m width~\cite{deblauwe2012determinants,gowda2018signatures}, while the earliest observations, from aerial photography, provide less than an 80-year window for tracking the pattern dynamics  predicted by models.   
Consequently, any proposal for using remote sensing data to validate and improve upon model predictions of dynamics is fraught with challenges.

In the absence of temporal data over an appropriate timescale, there has been a focus in some modeling efforts on predicting the dependence of spacing between vegetation bands as a function of climate characteristics such as mean annual precipitation or aridity, or trends with the elevation grades.  This, however, has proven a difficult test for validating models due to the predicted  multistability of patterned states,  and a lack of marked trends in observational data~\cite{sherratt2013history,siteur2014beyond}. Moreover, there appears to be a rather narrow interval of elevation grades that can support the banded patterns, e.g. typically $0.2\%-0.8\%$ \cite{deblauwe2011environmental}, which further limits any effort to validate trends suggested by model simulations.

One  direction of potential improvement for modeling efforts is to broaden predictions to include less utilized but still measurable spatial features of the patterns, such as the width and length of the bands (rather than only their spacing), and the curvature of the bands.
Such pattern characteristics may vary across a given site in response to locally-varying exogenous factors such as topography, soil type, and grazing or human pressure.    
Studying the response of the vegetation patterns to local changes in environmental conditions at a given site with uniform climate characteristics provides additional information about the system. 
The observed spatial variation may then be exploited to test and constrain models, rather than being treated as noise to be averaged over. Critical to such an effort is to have spatial information about the exogenous factor(s). 
The work in this manuscript is based on spatial variability associated with large scale topographic heterogeneity, which we study via globally-available digital elevation data sampled at a scale appropriate for pattern variation, e.g., $\sim 30$ m \cite{Farr:2007ib}.

The leading order effect of topographic heterogeneity on vegetation pattern formation is associated with the prevailing elevation gradient. This plays an  essential role in the formation and morphology of dryland vegetation patterns.  A grade as small as $\sim 0.2\%$ introduces anisotropy in the overland waterflow, thus favoring ordered banded patterns over isotropic `flat terrain patterns'~\cite{valentin1999soil,deblauwe2011environmental} Vegetation bands,  such as those of Fig.~\ref{fig:intro}, alternate rhythmically with stripes of bare soil, and are oriented perpendicular to the prevailing elevation grade.  
The earliest reported observations of vegetation bands are from the Horn of  Africa, and they  note the importance of even slight variation of the elevation grade in shaping banded vegetation patterns. 
The first such report, by MacFadyen in 1950~\cite{macfadyen1950vegetation}, points out that vegetation bands appear within modest topographic depressions or `valleys' aligned along the grade and are arced convex-upslope (Fig.~\ref{fig:intro}a). The subsequent 1957 observations reported by Greenwood~\cite{greenwood1957development}, provide additional examples, including  one where the bands exhibit a general convex-downslope pattern on a `ridge' between valley lines. The bands were even conjectured by Boaler and Hodge in 1964 to approximate  elevation contours~\cite{boaler1964observations}.

The arcuation of vegetation bands has remained largely unexplored in the modeling literature beyond the original efforts of Lefever and Lejeune~\cite{lefever1997origin,lejeune1999short}.  Simulations of their model, posed for a uniformly sloped terrain, show an ordered rectangular array of convex-upslope arced band segments. In this case the arcing is associated with a spontaneous break-up of a stripe pattern into a regular configuration of segments of the stripes that form the arcs, a possible deformation of the flat terrain
gap or spot patterns.
A similar phenomenon of spontaneous breakup and arcing of  the vegetation stripes was also reported for the generalized Klausmeier model on a uniform slope~\cite{siero2015striped}, in this case due to secondary modulational instabilities of stripes that occur as the precipitation parameter of the model decreases.  None of these modeling studies report observations of vegetation segments that are arced convex-downslope, nor do they investigate the type of topographic confinement of the patterns illustrated in Fig.~\ref{fig:intro}(b).

In this manuscript, we present a mechanism for arcing motivated by early observations of topographic influence on the shape of banded vegetation patterns and recent work highlighting the potential connection to surface water flow~\cite{iams2018topographic}. 
We employ a modeling framework within the class of reaction-advection-diffusion models that describes nonlinear interactions between continuous biomass and water fields, as well as advective water transport and diffusive biomass dispersal~\cite{meron2015nonlinear}.  Our investigations are based on the simplest and earliest model of this type, one proposed by Klausemeier~\cite{klausmeier1999regular} for striped patterns on a uniform hillslope. We generalize the water transport term in the model, denoted $T(W)$ below, to take into account possible positive and negative terrain curvature associated with valleys and ridges. We investigate the nondimensional form of the model, given by
\begin{align}
W_t &= T(W) - W - WB^2 + a \label{eq:klaus:W}\\
B_t &= \nabla^2 B - m B +WB^2,\label{eq:klaus:B}
\end{align}
where $W(x,y,t)$ is the water field and $B(x,y,t)$ is the biomass field.    Here the parameter
$a$ represents water input to the system in terms of the mean annual precipitation, and $m$ parametrizes the plant mortality rate. Water loss due to evaporation, in the dimensionless form of the equations, is captured by the $-W$ term and transpiration by the nonlinear $-WB^2$ term, which accounts for the biomass growth term $+WB^2$. The nonlinear dependence of the transpiration on biomass captures, in a heuristic fashion, a positive infiltration feedback between the biomass and the water.
In Klausmeier's formulation, the water transport is uniformly advected downhill
\begin{equation}\label{eq:transWklaus}
T(W)=T^K_{vx}(W) \equiv v W_x.  
\end{equation}
This term is essential for the existence of a family of stable traveling stripe patterns that migrate uphill~\cite{sherratt2005analysis,carter2018traveling}.

We consider a topographic extension of the Klausmeier water transport term, which is based on an assumption that water flow is proportional to the local elevation gradient. In dimensionless form, it  is given by 
\begin{equation}\label{eq:transW}
T(W)=T_{\zeta}(W) \equiv \nabla\cdot( W \nabla \zeta), 
\end{equation}
where $z=\zeta(x,y)$ is the elevation function.  See Table~\ref{tab:transW} for comparison of this topographic extension to other water transport models.
We use a terrain with simple channel-like geometry
\begin{equation}\label{eq:modelev}
\zeta=v(x+\sigma \cos(k_0 y))
\end{equation}
to explore the influence that the resulting modified water transport has on the predicted vegetation patterns. This idealized terrain consists of alternating ridge lines ($k_0 y=2\pi n$) and valley lines ($k_0 y=\pi +2\pi n$) 
aligned along a hillslope. 
This idealized elevation profile allows us to investigate how cross-slope terrain curvature, controlled by the parameter $\sigma$, influences the  vegetation patterns that form on ridges and in valleys.

More mechanistically detailed models than Klausmeier's, such as the Rietkerk model~\cite{rietkerk2002self} and Gilad model~\cite{gilad2004ecosystem}, differentiate between water in soil and surface compartments. This allows an infiltration feedback, a mechanism relevant for vegetation pattern formation in many environments, to be modeled explicitly.
The Gilad model~\cite{gilad2004ecosystem,meron2007localized} has also been formulated with a general topography for the surface water advection. It assumes an advective velocity in the transport term that is proportional to  $\nabla (\zeta+H)$,  where $H$ is the surface water height above the terrain. 
Other model studies have additionally sought to capture the influence of vegetation patterns in the formation of terracing of the underlying terrain along the hillslope through a coupling with erosion and soil transport dynamics~\cite{saco2007eco}.  The influence of topographic variation  at the scale of individual plants
has also been investigated in connection to the transition from banded vegetation patterns to vegetation following irregular drainage patterns~\cite{mcgrath2012microtopography}.   

Our investigation, based on our topographically-extended version of the Klausmeier model, leads to insight on the possible impact of 
pattern-scale ($\sim$ 100 m -- 1 km) spatial heterogeneity in topography on the shape of bands, and their location relative to the landscape. In particular, it leads to a simple interpretation of the possible impact of terrain curvature as increasing (for ridgelines) or decreasing  (for valleylines) water loss rate. The following prediction then emerges:  as aridity increases, the vegetation bands move from being located on ridgelines, convex downslope, to being confined to valleylines, convex upslope.

\begin{table*}[]
\centering
\caption{A summary of water transport models with references and equation numbers where appropriate. Here $W$ represents the water field in the Klausmeier modeling framework (Eq.~\eqref{eq:klaus:W}), $v$ characterizes the downhill waterflow rate on a uniform slope, $\gamma\ge 1$ allows for possibly nonlinear diffusion with rate $d$~\cite{van2013rise}, $H$ represents surface water height of the Gilad model~\cite{gilad2004ecosystem} and  $\zeta$ represents the elevation.   }
\label{tab:transW}
\begin{tabular}{l|rlcc}
Water transport model & Expression & & Eq. & Reference \\ \hline \\
 Original Klausmeier  & $T^K_{vx}(W)$ & $= v W_x$ &  \eqref{eq:transWklaus} &\cite{klausmeier1999regular} \\
 (Diffusion) generalized& $T^D_{vx}(W)$ & $=d\nabla^2(W^\gamma)+vW_x$ & - & \cite{van2013rise} \\ 
 Advection-only extension  &  $T^A_\zeta(W)$ & $=\nabla \zeta \cdot \nabla W$ &\eqref{eq:transWnaive} & - \\
 Topographic extension   & $T_{\zeta}(W) $ & $= \nabla\cdot( W \nabla \zeta)$ &\eqref{eq:transW} & -\\
&  &$= \nabla\zeta \cdot \nabla W  + W\nabla^2 \zeta  $ & \eqref{eq:transWadac} & \\
 Gilad &  $T^G_\zeta (H)$ & $=\nabla \cdot (H\nabla (\zeta+H))$ & - &\cite{gilad2004ecosystem} \\
   & &$= \nabla\cdot(H\nabla\zeta)  + \nabla^2 (H^2/2)  $ & - &
\end{tabular}
\end{table*}

This manuscript is structured as follows. In Sec.~\ref{sec:data}, we report the empirical relationship between arcing-direction of vegetation bands and the sign of the underlying terrain curvature for five sites in Western Australia, finding that convex-upslope arcs tend to occur within valleys, and convex-downslope arcs occur atop ridges.
In Sec.~\ref{sec:model} we investigate the influence of the proposed modification of the transport $T(W)$, Eq.~\eqref{eq:transW}, in the Klausmeier model for the idealized terrain given by Eq.~\eqref{eq:modelev}. 
We find that the topographically-extended Klausmeier model qualitatively captures key topography-dependent features such as the curvature of the arced patterns, in a manner consistent with our reported observations. 
Moreover it provides a way to characterize how terrain curvature impacts the location of vegetation patterns relative to both aridity and topographic features such as high vs. low ground in a gently rolling landscape. 
Finally in Sec.~\ref{sec:discussion}, we discuss how these predictions are borne out for sites on three different continents where patterns are observed. 
Additional details about simulations for the topographically extended Klausmeier model along with results based on a simplified version of the Gilad model with idealized terrain~\eqref{eq:modelev} appear in Appendices.

\section{Topographic influence on vegetation arcing in Australia}
\label{sec:data}

\begin{figure*}
\centering
\includegraphics[width=\textwidth]{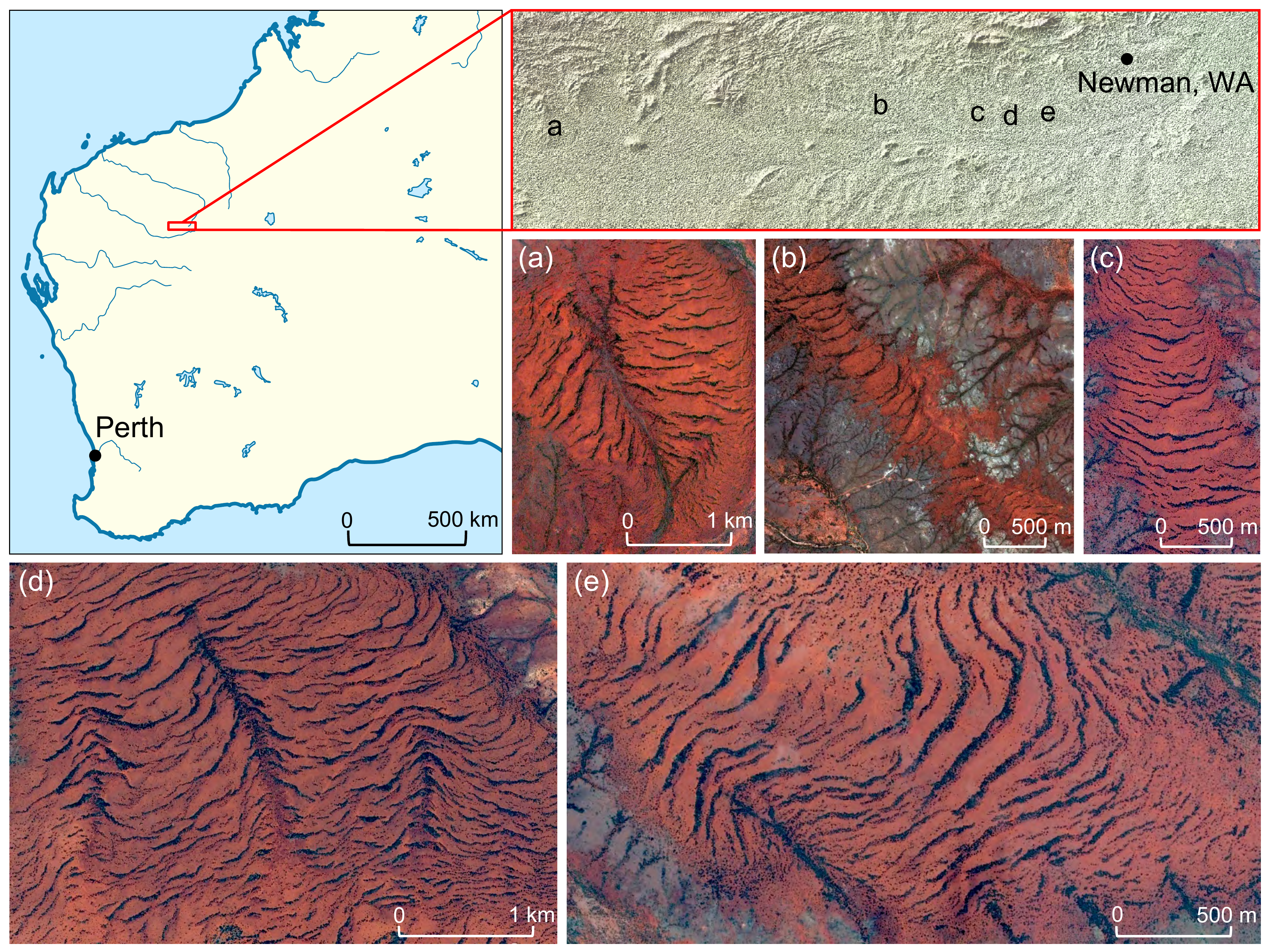}
\caption{  \label{fig:australia} 
Five sites displaying banded vegetation patterns in the Western Creek Basin southwest of Newman, Australia (\ang{-23.5} N, \ang{119.5} E), alongside a map indicating the relative location of each site.  The Sentinel-2A images~\cite{DRUSCH201225} were taken near the end of the Australian wet season, see scale bars for relative sizes. Map derived from image available via Wikimedia Commons.}
\end{figure*}

We  investigate the relationship between vegetation band arcing and terrain curvature in the Western Creek basin  (Fig.~\ref{fig:australia}) southwest of Newman, WA, Australia (\ang{-23.5} N, \ang{119.5} E), where vegetation bands have been previously reported~\cite{mcgrath2012microtopography}. A visual inspection of the area yielded 21 spatially-distinct sites featuring vegetation banding. Of these we selected the five with the most well-defined vegetation bands (Fig.~\ref{fig:australia}), ranging in area from 5-16 km$^2$.

We utilize two sets of remote sensing data in our analysis: Sentinel-2A multi-band satellite imagery~\cite{DRUSCH201225} with 10 m resolution, and a smoothed digital elevation model (DEM) provided by Geoscience Australia~\cite{gallant2011adaptive} with one arc-second ($\sim$ 30 m) resolution. Selected Sentinel images were taken on March 5, 2017, near the end of the Australian wet season and the topographic data was derived from the Shuttle Radar Topography Mission elevation dataset~\cite{Farr:2007ib}. 

\begin{figure*}
\includegraphics[width=\textwidth]{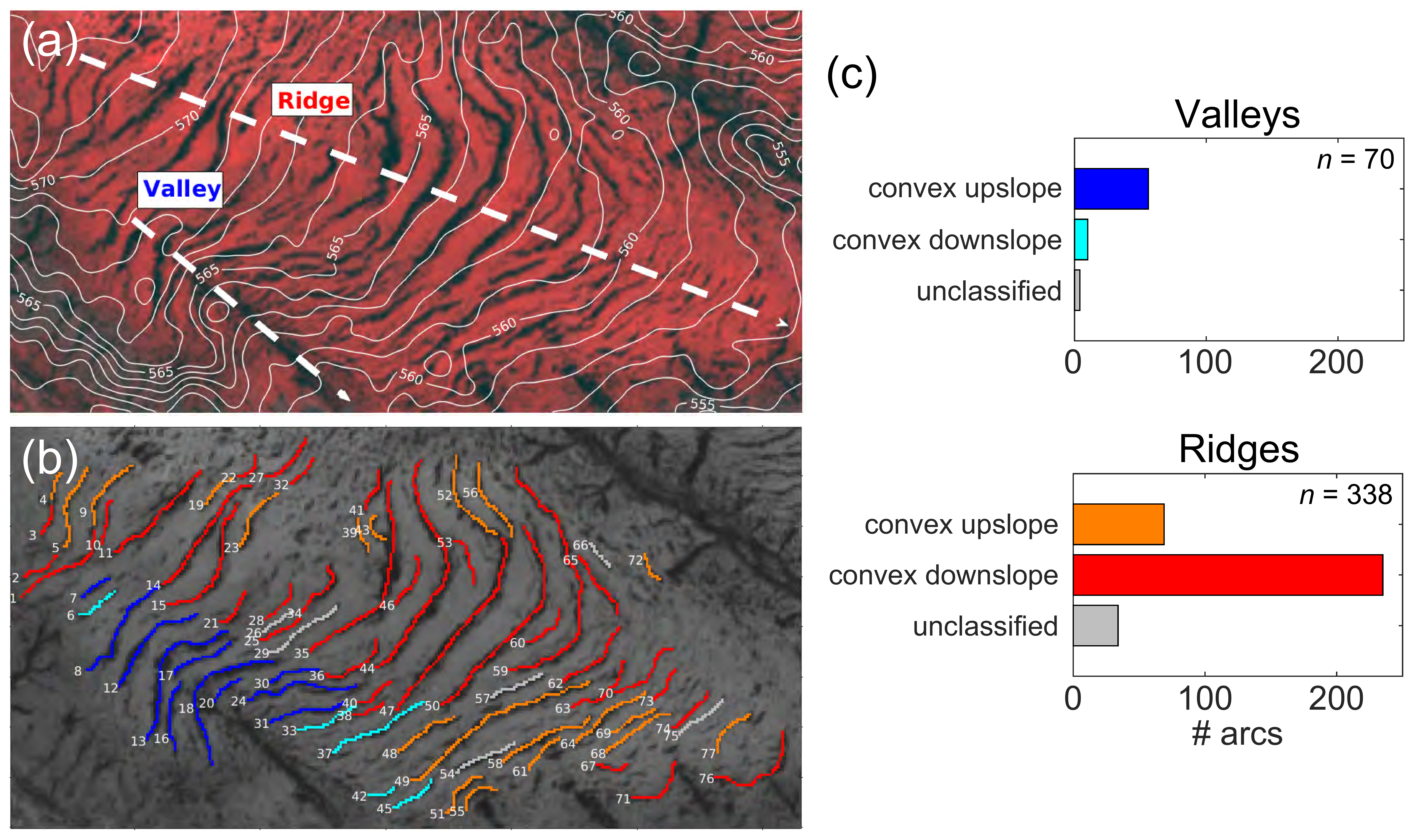}
\caption{  \label{fig:barplot} (a) An example site with elevation contours (solid white) and manually identified ridge/valley lines (dashed white, with arrows indicating the downhill direction). (b) The manually identified arcs are colored by the classification: convex-upslope in valley (blue), convex-downslope in valley (cyan), convex-upslope on ridge (orange) and convex-downslope on ridge (red). (c) Histogram of the 408 vegetation arcs from the five sites of Fig.~\ref{fig:australia}  that have been manually identified and classified as appearing in either a valley or on a ridge and as convex-uplsope or convex-downslope.   
}
\end{figure*}

At each site, we identify topographic ridge and valley lines by visual inspection of the elevation contour map (Fig.~\ref{fig:barplot}(a)). We then manually mark each identifiable vegetation band in a satellite image of the given site and record whether the band occurs in a valley or on a ridge (Fig.~\ref{fig:barplot}(b)). In instances where bands cross multiple ridge or valley regions, we divide the band into sections that are each restricted to a single valley or ridge.  We find that 82\% of the 408 bands identified across the five sites appear on ridges.  For vegetation bands with a visually-identifiable direction of curvature (91\% of bands), we further classify them as arcing in the convex-upslope or convex-downslope directions (Fig.~\ref{fig:barplot}(b,c)). We identify 80\% of the bands within valleys  as convex-upslope and 70\% of the bands on ridges as convex-downslope.

This evidence supports the observational claim that the direction of arcing of the vegetation bands correlates with the sign of curvature of the underlying terrain~\cite{macfadyen1950vegetation, greenwood1957development, boaler1964observations}.  We note that our analysis does not discern a curvature direction for 9\% of the bands, and topographic features that appear on length scales of a few hundred meters or less (e.g., the small scale ridges and valleys at the bottom of Fig.~\ref{fig:barplot}(a,b)) were neglected.  This may account for some of the inconsistencies between the data and observational claim.

\section{Modeling the influence of topography}\label{sec:model}

The majority of vegetation bands at the Australian sites (Fig.~\ref{fig:australia}) appear on ridges and the bands in the Horn of Africa (Fig.~\ref{fig:intro}) tend to appear in valleys, and observations across both sites suggest a consistent trend that vegetation bands arc convex-upslope in valleys and convex-downslope on ridges.   In this section we show, using the modeling framework proposed by Klausmeier as a template, that this topographic influence can be captured via water transport.  In particular we motivate and explore a \textit{topographically-extended Klausmeier model}   consisting of Eqs.~\eqref{eq:klaus:W}-\eqref{eq:klaus:B} with water transport $T_\zeta(W)$ given by Eq.~\eqref{eq:transW} on the idealized terrain with elevation $\zeta$ given by Eq.~\eqref{eq:modelev}. 
We also extract predictions from this model about how the location of patterns for a given terrain, e.g. whether  patterns are predominantly on ridges or valleys, may depend on the level of aridity.

The topographically-extended model reduces to the original Klausmeier model on uniformly sloped terrain or the original Klausmeier model with a modified linear water loss rate on uniformly curved terrain.  We can express Eq.~(\ref{eq:transW}) as   
\begin{equation}\label{eq:transWadac}
T_\zeta(W)= \nabla \zeta\cdot\nabla W +  (\nabla^2 \zeta) W
\end{equation}
and note that  the first term on the right-hand side represents advection downhill with rate proportional to the local slope and the second term represents accumulation of water if the terrain curvature is positive and diversion of water if the curvature is negative. In the case that $\nabla^2\zeta$ is constant, this uniform curvature term adds a term proportional to $W$ into Eq.~(\ref{eq:klaus:W}). In Sec.~\ref{sec:model:effevap}
we  exploit this observation in our analysis to predict when patterns might be expected to form on ridges versus valleys within the model.

\subsection{A model topography}
\begin{figure*}
\centering
\includegraphics[width=.9\textwidth]{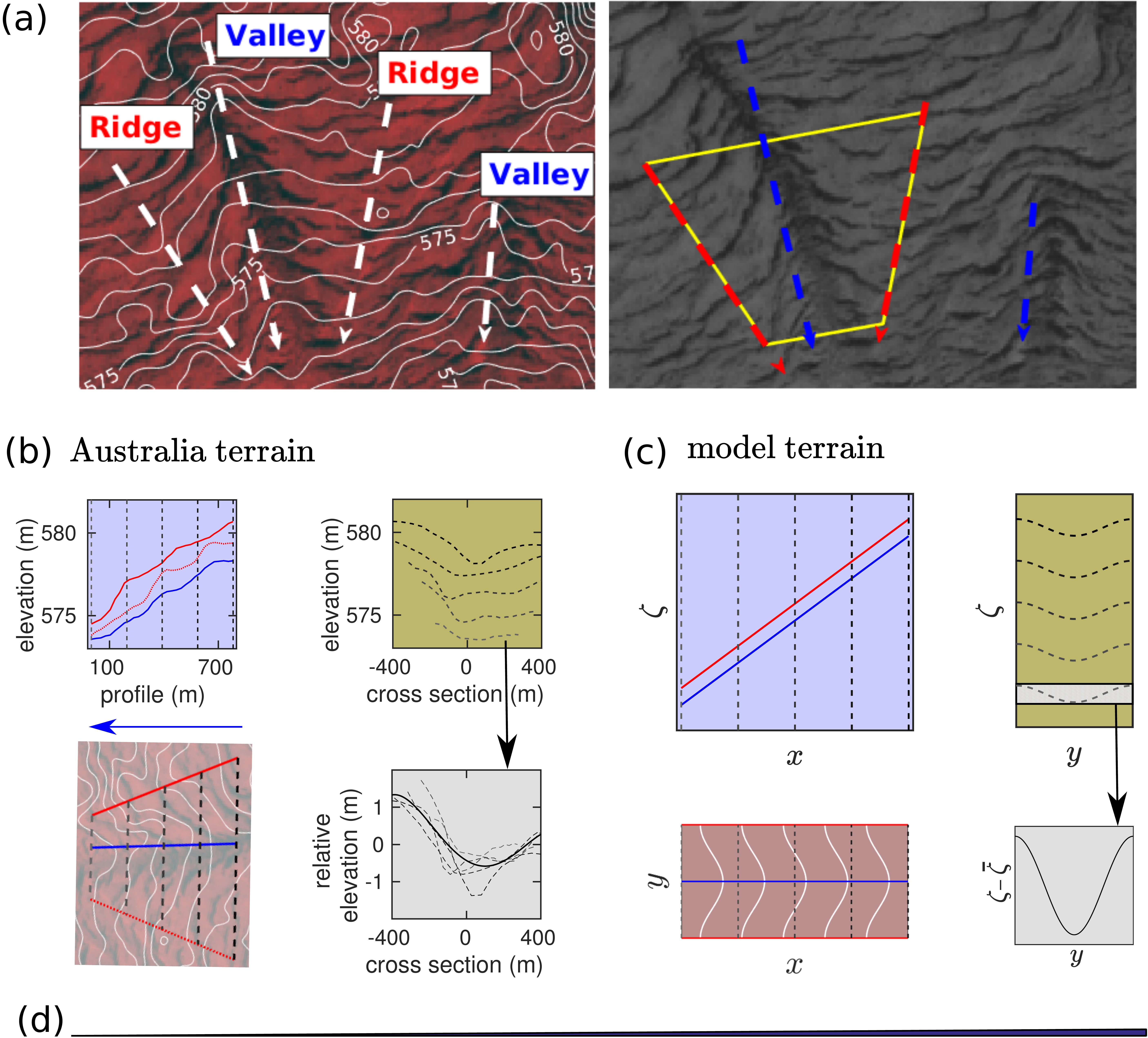}
\caption{  \label{fig:topo}  (a) The left image shows elevation contours (solid white) on a section of the patterned site in Fig.~\ref{fig:australia}(d). A few approximate ridge and valley lines (white dashed) are shown with arrows pointing downhill.  The yellow trapezoid in the right grayscale image indicates a typical topographic structure consisting of a valley aligned along the grade surrounded by two ridges.  (b) Elevation contours for the trapezoid region in (a) are shown together with their profiles along the grade (above) and along cross sections transverse to the grade (right). A fourth order polynomial fit (solid black) is shown for the elevation relative to the mean of the cross sections (dotted gray).  Note that the profile distance is taken to be distance along the valley line and the associated ridge elevations are found by orthogonal projection.  (c) Model topography given by Eq.~\eqref{eq:modelev} with $v=10$, $\sigma=5$, $L_x=100$ and $L_y=50$.  Elevation contours along with cross sections shows the height of ridge (red) and valley (blue) relative to the change in elevation that results from uniform slope along $x$.  Note that the scale for elevation in (a) and (b) is greatly exaggerated relative to the $x$ and $y$ dimensions. (d) For reference, the triangle at the bottom is drawn to scale with a 0.6 \% grade corresponding to the mean of the example shown in (b). }
\end{figure*}

Figure~\ref{fig:topo}(a) shows an example of banded vegetation patterns organized around modestly curved topographic `valleys' and `ridges' aligned along the slope at the Australia site in Fig.~\ref{fig:australia}(b). Figure~\ref{fig:topo}(b) provides detailed topographic information about the trapezoid region marked in Fig.~\ref{fig:topo}(a).  
The valley line follows an approximate 0.6\%  average grade downhill, and a fourth-order polynomial fit of elevation relative to the mean for five equally-spaced cross sections show the typical cross sectional ridge-valley elevation difference is on the order of one meter, while the distance between the ridge lines is more than 300 meters. 

As a simple approximation of this characteristic topographic variation, we consider the idealized topography given by Eq.~\eqref{eq:modelev} and depicted in Fig.~\ref{fig:topo}(c), consisting of a periodic array of valleys and ridges aligned along the $x$-axis with uphill in the positive $x$-direction. 
In our model simulations,  we fix the dimensionless slope parameter to $v=10$, similar to the value in~\cite{gowda2018signatures}. 
Based on mesh size considerations we choose 
 $k_0=2\pi/L_y$ with $L_y=50$.
The parameter $\sigma$, which we refer to as the channel `aspect', controls the cross-slope 
elevation difference between the ridge and valley lines.
We consider channel aspect $ 0 \le\sigma \le 6$; the average slope transverse to a channel from a valley to a ridge is smaller, but of the same order of magnitude, as the slope along the $x$-direction, consistent with Fig.~\ref{fig:topo}(b).  The upper bound on the channel aspect parameter $\sigma$ is restricted by our choice of water transport, Eq.~\eqref{eq:transW}, and terrain parameters. For $\sigma>1/vk_0^2 \approx 6.33$ the effective water loss rate in Eq.~\eqref{eq:klaus:W} along the valley line $y=L_y/2$ will be negative, causing a  global and potentially unphysical change in the solution structure of the model.

\subsection{Simulation results}
We use the idealized terrain given by Eq.~\eqref{eq:modelev}, fix the plant mortality to $m=0.45$~\cite{klausmeier1999regular}, and explore the influence of precipitation $a$ and channel aspect $\sigma$ on patterns within the topographically-extended Klausmeier model. Time simulations were carried out using a fourth order exponential time differencing scheme~\cite{cox2002exponential,kassam2005fourth}, more details and additional numerical results appear in App.~\ref{app:sim}.

For uniformly sloped terrain, the water transport is the same as Klausmeier's (i.e. $T_\zeta(W)=T_{vx}^K(W)$), and there exists a uniformly vegetated state, with $B=a/2m + \sqrt{(a/2m)^2-1}$, which is stable provided the precipitation level $a$ is sufficiently high. 
The uniform state loses stability if precipitation drops below some threshold $a=a_T$, a bifurcation that produces a periodic traveling wave pattern consisting of straight vegetation bands transverse to the slope that are migrating slowly uphill~\cite{sherratt2005analysis}. A lower threshold $a=a_{O}$ marks the smallest precipitation level that can support just a single isolated band of vegetation on the domain, the so-called `oasis state'~\cite{van2013rise}.   Within the range $a_{O}<a<a_T$, there exists a multiplicity of stable traveling patterns consisting of periodically-spaced bands, each with different periods, and possibly different orientations relative to the grade~\cite{siero2015striped}. When there is too little precipitation, i.e. $a<a_{O}$, only the bare soil or `desert state' exists.

\begin{figure}
\centering
\includegraphics[width=.75\columnwidth]{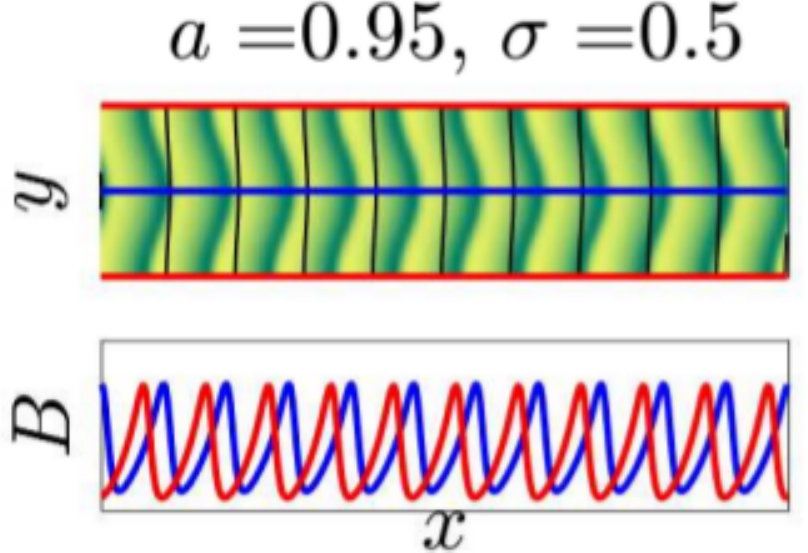}
\caption{  \label{fig:arcsim} Upper frame shows biomass field  on the full two-dimensional domain at $t=1000$ with the modestly curved elevation contours (black) superimposed.  Green (yellow) on the color scale represents high (low) biomass values.  While the  vegetation bands are more  significantly arced than the elevation contours, the direction of curvature consistently matches across the domain for both. The lower frame shows one-dimensional profiles of the biomass along ridge (red) and valley (blue) lines. Periodic domain: $L_x=200$, $L_y=50$. Parameters: $a=0.95$, $m=0.45$, $v=10$, $k_0=2\pi/L_y$. }
\end{figure}

For small channel aspect, e.g. $\sigma=0.5$ as in Fig.~\ref{fig:arcsim},  numerical simulation indicates that traveling band patterns persist and develop a transverse modulation such that the arcing-direction of the vegetation bands matches the direction of curvature of elevation contours.  We note that, while their signs of curvature match, the vegetation bands do not closely follow the elevation contours.  Moreover, in numerical experiments, where we turn off the  accumulation term $(\nabla^2\zeta) W$ of $T_\zeta(W)$ in Eq.~\eqref{eq:transWadac}, we find that the bands straighten out into the banded patterns of the original Klausmeier model. Hence the advective term alone does not account for the curvature of the bands in our simulations.

\begin{figure*}
\centering
\includegraphics[width=\textwidth]{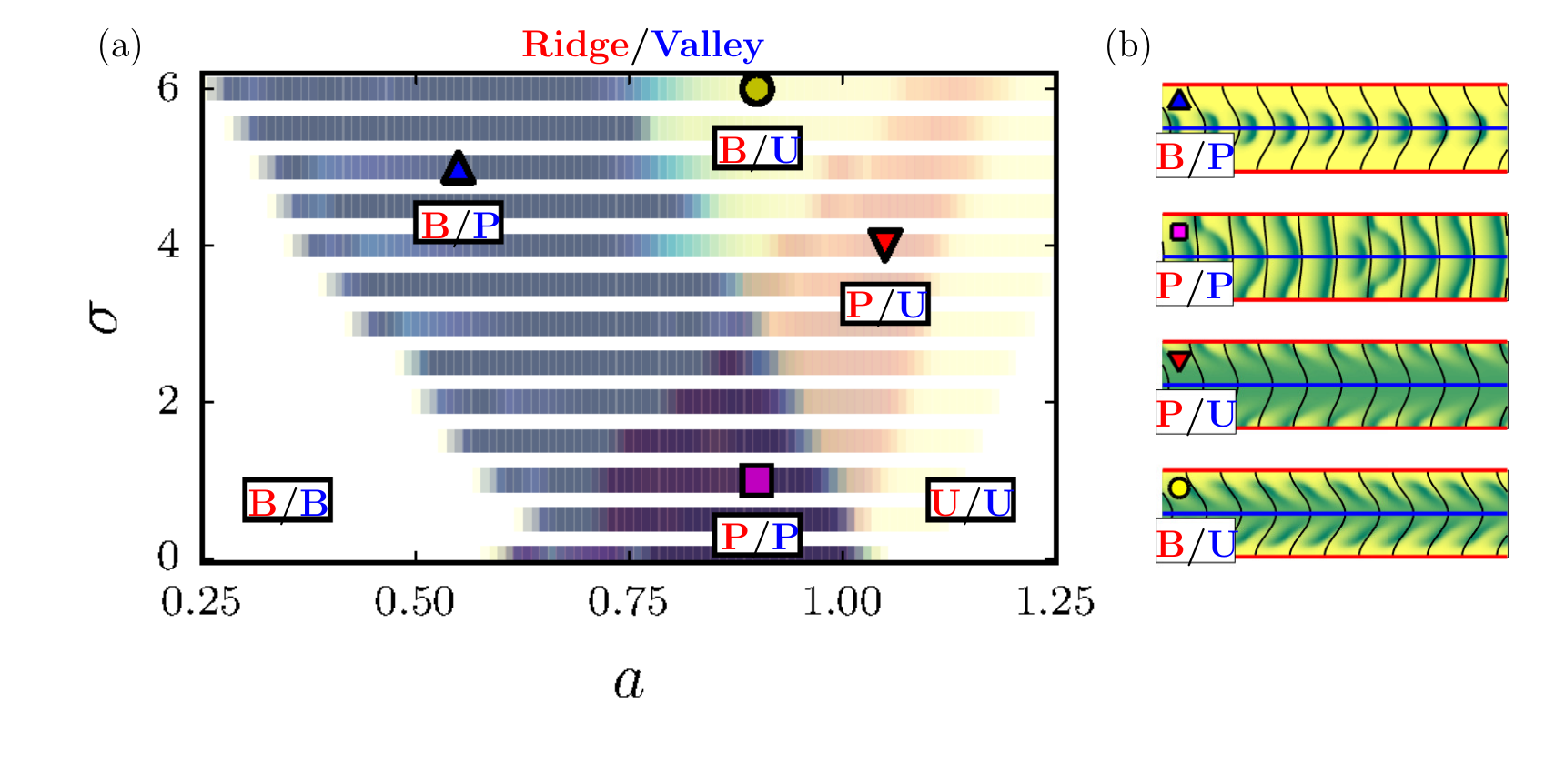}
\caption{  \label{fig:parspace}  (a) Maximum of amplitude of patterns  on ridges and valleys in $(a,\sigma)$-plane when $a$ is decreased and then increased for fixed $\sigma$.  Yellow indicates small amplitude while red (blue) indicates large amplitude on ridge (valley) and purple indicates large amplitude on ridge and in valley.   
Each region is labeled by ridge/valley state as bare soil (B), patterned (P) or uniformly vegetated (U).  (b) Biomass from simulations initialized with uniform vegetation and parameters associated with the labeled points the $(a,\sigma)$-plane are shown for $t=1000$. Parameters: $m=0.45$, $v=10$, $k_0=2\pi/50$. }
\end{figure*}

The bottom panel of Fig.~\ref{fig:arcsim} shows little difference between the pattern characteristics on the ridgeline and those along the valleyline. However, for deeper channels, associated with larger channel aspect $\sigma$, we find that the pattern characteristics
may differ significantly between  ridges  and  valleys, with the further possibility that patterns may end up confined to some part of the terrain. 
Figure~\ref{fig:parspace} summarizes the various types of patterns  in the $(a,\sigma)$ parameter plane, classified by their ridge/valley states, as either bare soil (B), patterned (P) or uniform vegetation (U).

Figure~\ref{fig:parspace}(a) shows the maximum ridge (valley) pattern amplitudes obtained from simulations by red (blue) intensity at each point in the $(a,\sigma)$-plane.  Purple indicates where patterns appear on both ridges and valleys while yellow indicates where no significant pattern amplitudes appear on either ridges or valleys. The visualization of the simulation data in Fig.~\ref{fig:parspace}(a) does not distinguish between uniform vegetation cover and bare soil so we additionally label each region of the parameter space based on the ridge/valley states (e.g. B/P indicates bare ridge and patterned valley).   Figure~\ref{fig:parspace}(b) provides example biomass profiles with ridge (valley) lines along which pattern amplitudes are computed marked in red (blue). Additional details about the simulations used to generate Fig.~\ref{fig:parspace} are provided in Appendix~\ref{app:sim}.

Considering the effects of water redistribution, in directions transverse to the slope,  can provide some insight into the emergence of the various types of patterns shown in Fig.~\ref{fig:parspace}. We highlight the following observations:
 
\begin{itemize}
\item{} For low water availability (e.g. blue up triangle) arced vegetation segments are confined to valleys (B/P). The topography diverts water away from the ridges which remain bare even though the precipitation level would be sufficient to sustain stripe patterns on a uniformly sloped domain. Simulations also show that these confined patterns persist to lower precipitation levels than would be possible on uniformly sloped terrain.  

\item{}For larger precipitation values (e.g. purple square), some of the arcs may connect across the ridges to form continuous bands (P/P).  This occurs for precipitation parameters where patterns would appear on uniformly sloped terrain. It also requires  shallow  topographic variation transverse to the grade, so that the water redistribution is insufficient to cause bare soil on ridges or uniform vegetation cover in  valleys.  

\item{} At higher precipitation levels  (e.g. red down triangle), patterns consist of uniform vegetation cover in valleys and arced gaps on ridges (P/U).  In this case water is diverted away from the ridges allowing the pattern to persist even when the precipitation is high enough to sustain a uniform vegetation on uniformly sloped terrain.  

\item{}If the aspect $\sigma$ is large enough (e.g. yellow circle), water diversion and accumulation will result in uniform vegetation in valleys and bare soil on ridges. We find in this case that the patterns may be confined,  chevron-like, in the intermediate topographic zone between the uniformly vegetated valleys and the bare ridges.
\end{itemize}

\subsection{Terrain curvature  as an effective change in water loss rate}
\label{sec:model:effevap}

The terrain acts to divert water from ridges to valleys, effectively altering local rates of water loss and accumulation. 
Focusing on ridge and valley lines allows us to use this interpretation to make predictions about the transitions between the various vegetation states of the two-dimensional, spatially inhomogeneous model based on two spatially homogeneous one-dimensional models with different water loss rates (Fig.~\ref{fig:parspace}).  Specifically, we consider one-dimensional models for ridge and valley lines of the form  
\begin{align}
W_t &= v W_x - r W - WB^2 + a \label{eq:klaus1d:W}\\
B_t &=  B_{xx} - m B +WB^2,\label{eq:klaus1d:B}
\end{align}
where the  parameter $r$ 
characterizes an effective water loss rate and it differs  between the ridges and the valleys.  Using the idealized topography given by Eq.~\eqref{eq:modelev}, the water equation, Eq.~\eqref{eq:klaus:W}, along a ridge or valley line reduces to Eq.~\eqref{eq:klaus1d:W} with effective water loss rate given by
 \begin{align}
r_R &=1-\nabla^2\zeta\big|_R=1+ v\sigma k_0^2\\ 
r_V &=1-\nabla^2\zeta\big|_V=1- v\sigma k_0^2. \nonumber
\end{align}
On ridges the effective water loss rate $r_R$ is increased by the fact that water is diverted away, and in  valleys the effective water loss rate $r_V$ is decreased by the accumulation of water. Neglecting the $B_{yy}$ term in the biomass equation, Eq.~\eqref{eq:klaus:B}, the patterns along ridge and valley lines can each be described by the system of one-dimensional PDE's \eqref{eq:klaus1d:W}-\eqref{eq:klaus1d:B} that correspond to Klausmeier's original model with an effective water loss rate $r$.   As with the two-dimensional version of the model, we restrict our parameters to ensure that the effective water loss rate always remains positive: $\sigma \le 1/v k_0^2$. 

In the context of this one-dimensional approximation the uniform state loses stability, leading to the onset of pattern formation with decreasing precipitation, in a so-called Turing-Hopf  bifurcation (TH) at $a_T$. Weakly nonlinear analysis shows that the periodic traveling wave solution branch emanating from TH will be supercritical for all values of $\sigma$ in valleys and for $\sigma \lesssim 0.85$ on ridges.  In the cases where it is subcritical, numerical continuation shows that the interval $\Delta a$ of coexistence of patterned states and the uniform state for fixed $\sigma$ is typically $\lesssim 3\%$ of the interval of existence for patterned states.  We can therefore reasonably approximate the upper bound for the existence of patterns by $a_T$.  The disappearance of the oasis state, consisting of a single pulse of vegetation, occurs through a saddle-node bifurcation (SN$_O$) at $a_O$.  Numerical continuation of the one-dimensional model shows that, for fixed $\sigma$, no other patterned state that bifurcates from the uniform state exists at lower values of $a$. We therefore take $a_O$ as the lower bound of the range of existence of patterns.  The interval for existence of patterns in the one-dimensional models is therefore given by $a_{O}<a<a_{T}$~\cite{van2013rise,sherratt2007nonlinear}. 

\begin{figure*}
\centering
\includegraphics[width=\textwidth]{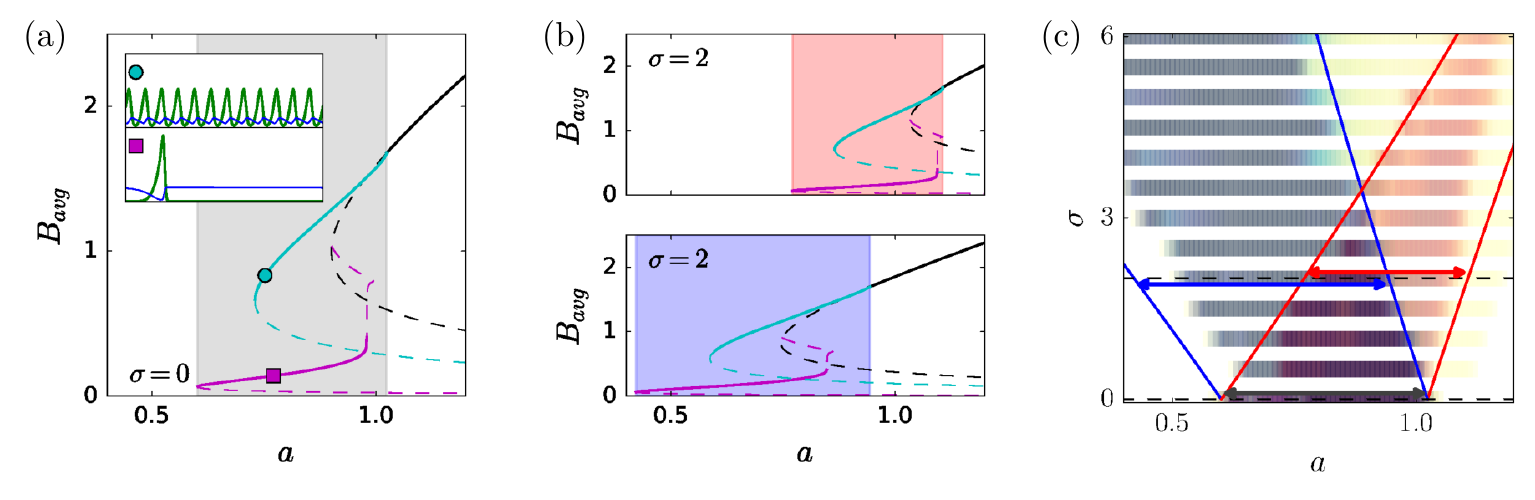}
\caption{  \label{fig:bif}  (a) Bifurcation diagram of one-dimensional Klausmeier model given by Eqns.~\eqref{eq:klaus1d:W}-\eqref{eq:klaus1d:B} showing uniform vegetation state (black) and two of the family of traveling wave solutions that bifurcate from the upper branch of this state: the Turing-Hopf branch (cyan) with wavelength approximately 16.27 and the branch corresponding to a single pulse of biomass on the domain $L_x=200$ (magenta). Stable (unstable) solutions are indicated by thick solid (thin dashed) lines and at least one solution from this family is stable within the interval of $a$ shaded in gray.  (b) Bifurcation diagrams for model with effective water loss rate corresponding to the ridge and  valley for $\sigma=2$ are compared to the model with $\sigma=0$.  The region shaded red (blue) indicates existence of stable traveling wave solutions for the model associated to the ridge (valley). (c) Boundaries of predicted regions of existence for ridge (red) and valley (blue) patterns based on one-dimensional models. The results are superimposed on the two-dimensional simulations shown in  Fig.~\ref{fig:parspace}(a) for comparison.   Parameters: $v=10$, $m=0.45$, $k_0=2\pi/50$.  These numerical continuation results were computed with AUTO~\cite{doedel2012auto}, see Sec.~\ref{sec:model:effevap} for more details.     
}
\end{figure*}

Figure~\ref{fig:bif}(a) shows the bifurcation diagram computed via numerical continuation~\cite{doedel2012auto} for the uniform vegetation branch, TH branch, and oasis branch for $\sigma=0$ along with examples of biomass profiles for these two patterned states. We omit other branches with intermediate numbers of vegetation stripes between these two extremes.  Since the ridge and valley models both reduce to the original Klausmeier model in one dimension for the $\sigma=0$ case, the gray shaded region indicates the interval of existence for patterns for both.  As $\sigma$ increases, illustrated by $\sigma=2$ in Fig.~\ref{fig:bif}(b),  the bifurcation structure of these states remains qualitatively the same but the interval of existence is shifted to higher precipitation values on ridges and lower precipitation values in valleys. The difference in behavior here reflects the intuition that more water is available to the vegetation in the valleys than on the ridges.

Figure~\ref{fig:bif}(c) shows the interval of existence for the 1-dimensional ridge (red) and valley (blue) patterns in the $(a,\sigma)$-plane, superimposed on the results of our two-dimensional simultions from Fig.~\ref{fig:parspace}(a). 
The lines, indicating birth and death of one-dimensional patterns on ridges and valleys, slightly overestimate the range of patterns in the ($a,\sigma)$-plane  compared to the full two-dimensional model. This may be due to early transitions between states when noise is added to the numerical simulations or because of interactions between the ridge and valley neglected by the one-dimensional models. 
We note that for sufficiently small channel aspect $\sigma$ the  existence intervals overlap, and  the model predicts  states with patterns both on ridges and in valleys.  Near $\sigma\approx 3.5$, there is a transition point above which the existence regions for ridge and valley patterns no longer overlap. In this parameter range, the prediction is that patterns may exist in the valleys, with bare ridges, for low precipitation $a$, or, for higher $a$, patterns may be confined to the ridges, with uniform vegetation in the valleys. 

\section{Ridge and valley patterns in Australia, Ethiopia and USA}\label{sec:discussion}

Our analysis of the topographically-extended Klausmeier model suggests that water transport resulting from terrain curvature plays an important role in generating arced vegetation bands. 
The model predicts that regions with relatively greater water availability (represented by the precipitation parameter $a$ in the model) will exhibit vegetation patterns on ridges where water is diverted away, and regions with relatively low water availability will exhibit patterns in valleys where water can accumulate. While we do not expect to be able to make quantitative predictions within Klausmeier's simple modeling framework, we can assess  qualitative predictions through observations of vegetation on ridges and in valleys from remote-sensing data.  We assume that a given site samples  different ridge and valley curvatures with approximately no change in water availability. We note that, in contrast to the the idealized terrain underlying the predictions summarized in Fig.~\ref{fig:parspace}, the height and width associated with adjacent ridges and valleys are rarely equal for real topographies.

Banded vegetation patterns are observed in dryland environments with mean annual precipitation (MAP) ranging from 150 to 800 mm~\cite{deblauwe2008global}. Factors such as temperature and soil characteristics can strongly influence the amount of water effectively available to plants for growth, so it is useful to characterize water availability by the aridity index (AI), which captures the degree of imbalance between the influx of rainfall and the potential outflux of water due to evaporation and plant transpiration~\cite{middleton1992world}. The sites in Australia considered in Sec.~\ref{sec:data} (Fig.~\ref{fig:australia}) have an aridity index of 0.15, falling near the middle of the range where patterns are typically observed (AI = 0.05 to 0.3)~\cite{deblauwe2008global}. We discuss the predictions of the topographically-extended Klausmeier model in the context of these Australia sites along with two additional sites: the less arid Ft. Stockton, USA (AI $=$ 0.19) and the more arid Haud region of Ethiopia (AI $=$ 0.08). 
Table~\ref{tab:sites} details locations and water availability characteristics for all three sites.

\begin{table*}[]
\centering
\caption{Banded vegetation sites considered in this study, along with aridity index (AI), potential evapotranspiration (PET), and mean annual precipitation (MAP) values for each site. Values for AI and PET are given by the Global Aridity and PET Database~\cite{ZOMER200867} and MAP is computed from the product of these values.   }
\label{tab:sites}
\begin{tabular}{r|lllllll}
Site 		&Continent  	&Coordinates  	&AI		&PET (mm) &MAP (mm) 	& References\\ \hline
Haud			&Africa  		&(8.00, 47.58) 	&0.08 &  1889	&144 		& \cite{macfadyen1950vegetation,greenwood1957development,boaler1964observations} \\
Newman	 		&Australia  	&(-23.50,119.60)	&0.15  & 1881	&278  		& \cite{mcgrath2012microtopography}\\
Ft. Stockton 	&North America  &(31.60,-103.08) &0.19  &	1676 &320  		& \cite{mcdonald2009ecohydrological,penny2013local}
\end{tabular}
\end{table*}

As noted  in Sec.~\ref{sec:data}, the majority of vegetation bands at the Australia sites appear on ridges which tend to be oriented convex-downslope.  The valleys are largely populated by uniform vegetation cover, and the vegetation bands that do appear within valleys tend to be oriented convex-upslope. In the context of the topographically-extended model, the Australia sites therefore appear to fall above the transition point (e.g., $a\approx 0.9$ in Fig.~\ref{fig:parspace}) where a wider range of curvatures support ridge patterns than valley patterns but below the upper bound (e.g., $a\approx 1$ in Fig.~\ref{fig:parspace}) for the existence of valley patterns at any curvature value.

\begin{figure*}
\includegraphics[width=\textwidth]{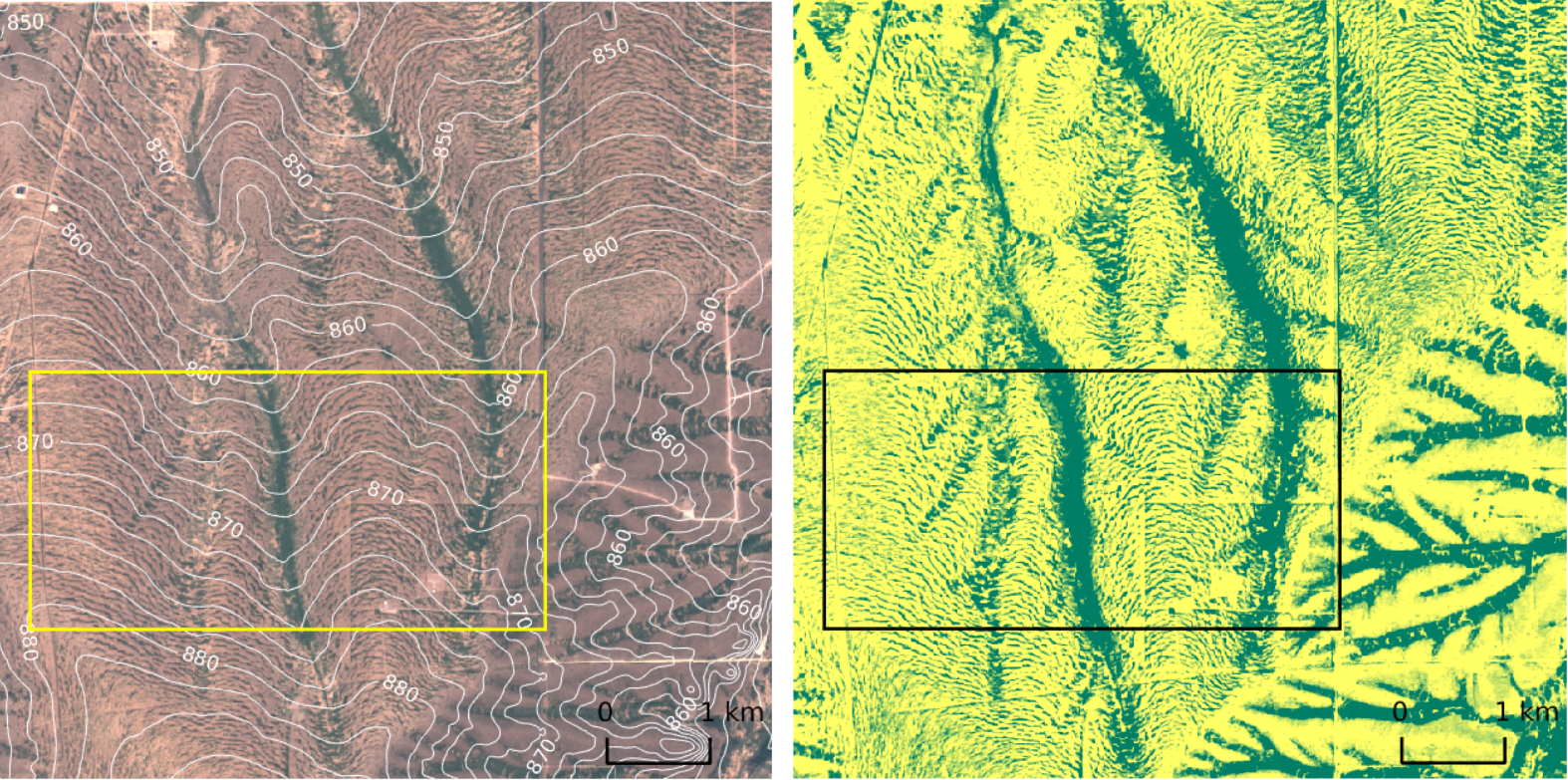}
\caption{ \label{fig:FtStock} 
Satellite image (left) and maximum NDVI (right) of a 7 km by 7 km region Northwest of Ft. Stockton, TX, USA that exhibits vegetation bands on the Northwest face of a hill. The boxed region shows a 5 km by 2.5 km area with a series of ridges and valleys aligned along the slope. The broad ridge on the right of the box is about 2 km across, measured between the two uniform vegetation valleys,  and has a valley-to-ridge elevation gain of about 5 m.   
Four smaller valleys can also be observed with characteristic transverse width of 300-400 m and transverse elevation changes less than a meter.    The mean slope along this NW face is about 0.6 \% ($\sim$40 meters over 6.6 km).   The Southeast face is steeper, with an average grade of about 0.75 \%, and the vegetation is organized into a network of channels with ridge-to-ridge spacing ranging from 300 m to 800 m of and valley to ridge depth ranging from 1 m for the narrower channels to 6 m for the larger central channels that the narrower channels feed into.  The satellite imagery comes from Sentinel 2~\cite{DRUSCH201225} while the elevation comes from the US National Elevation Dataset~\cite{gesch2002national}.  The RGB image was chosen from the least cloudy day between October 2016 and May 2017. High (low) values for maximum NDVI at each pixel over the two year period May 2015 to May 2017 are shown with green (yellow). }
\end{figure*}

The region near Ft. Stockton, TX, USA~\cite{mcdonald2009ecohydrological,penny2013local} shown in Fig.~\ref{fig:FtStock} exhibits vegetation bands on the northwest face of a gentle slope, while the southwest face is populated by vegetation organized within what appears to be a network of drainage channels.  
The bands appear mostly on ridges (oriented convex-downslope) but also in shallow valleys (oriented convex-upslope), as expected by the relatively high aridity index at this site.  The series of ridges and valleys within the box displays uniform vegetation cover in the two deeper valleys and patterns in the shallower valleys, with the valley patterns more densely vegetated than the ridge patterns. Further analysis is required to understand the extent to which the topographically-extended model captures the observed differences in morphology between the northwest and southeast faces, as well as the extent to which other factors such as slope, aspect or microtopography play a role.

\begin{figure*}
\centering
\includegraphics[width=0.95\textwidth]{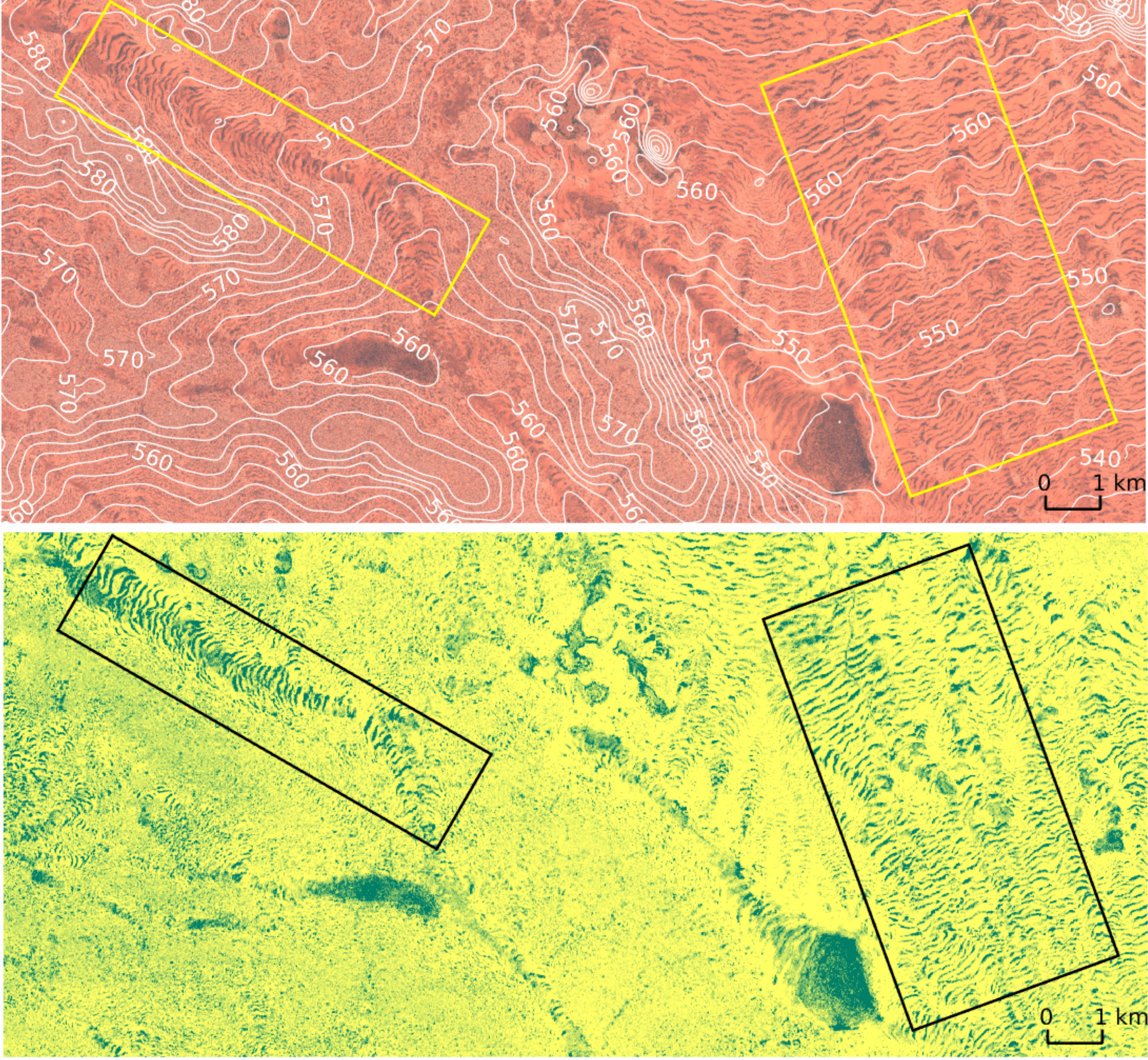}
\caption{  \label{fig:Eth} 
Satellite image (upper) and maximum NDVI (lower) of a 21 km  by 10 km region of Ethiopia located just east of the site shown in Fig.~\ref{fig:intro}(b).  
Two areas within the image are highlighted by boxes.  The left is a channel aligned along a overall slope of about 0.2\% grade with an approximate ridge-to-ridge distance of 3 km.  The vegetation pattern, with bands arcing convex-upslope, appears within a 2 km region surrounding the valley line where the transverse elevation has a 2 meter variation from minimum to maximum. The grade transverse to the slope increases to about 1\% on the southwest side of the valley while remaining roughly at 0.1\% on the northeast side.  The region highlighted on the right is within a broad, relatively flat valley with an overall 0.3\% grade and transverse variations of less than 1 m across the 4 km width of the box.  
The satellite imagery comes from Sentinel 2~\cite{DRUSCH201225} while the elevation data is ALOS~\cite{tadono2014precise} with spectral smoothing~\cite{shirai2014fft}.  The RGB image is of the least cloudy day between October 2016 and May 2017 and the NVDI shows maximum value at each pixel over the same time period. }
\end{figure*}

The region in Ethiopia shown in Fig.~\ref{fig:Eth} is located just east of the site shown in Fig.~\ref{fig:intro}(b). As expected by the relatively low aridity index, vegetation bands are confined to valleys, with sparse vegetation cover on ridges.  The elevation contours indicate that the ridges are more sharply curved than the valleys, and the vegetation tends to appear where the grade is relatively low ($\sim$0.2-0.3\%).   The boxed area on the left exhibits banded vegetation, clearly oriented convex-uplsope, within a relatively narrow valley.  The boxed area on the right is situated within a broad, relatively flat region. Here, there is neither a consistent trend in the direction of vegetation band arcing, with bands often appearing nearly straight, nor is there a consistent trend in the curvature of the topography. Elevation data with higher spatial resolution may help address whether the processes that set the observed lengths and curvatures of the vegetation arcs in broad flat regions of this type are driven by  variation in topography or by other effects.

In addition to vegetation patterns, the region shown in Fig.~\ref{fig:Eth} exhibits two large basins with uniform vegetation cover.  While the curvature transverse to the grade is not significantly greater in the basins than in the patterned valleys nearby, these  basins are formed from curvature along directions both transverse and parallel to the slope. Since the topographically-extended model predicts that both curvatures will contribute to the effective decrease in water loss rate, uniform vegetation cover within basins of this type is qualitatively consistent with the model predictions.

\section{Conclusion} 

We have developed an extension to the Klausmeier model for banded vegetation patterns that captures the influence of terrain curvature on water transport. The model simulations indicate a correlation between the arcing-direction of vegetation bands and the sign of curvature of the underlying terrain, consistent with observational evidence from sites in Australia presented in Sec.~\ref{sec:data}, and also early observational studies of patterns in the Horn of Africa~\cite{macfadyen1950vegetation,greenwood1957development,boaler1964observations}. 
While the simulated vegetation bands arc in the same direction as the underlying elevation contours in the topographically-extended model, they do not always closely match the contours. The model therefore predicts that the bands do not sit transverse to the local gradient everywhere, which
has seen some support in observations relating to small-scale ($\sim$ 10 m) disruptions of water flow~\cite{penny2013local}. Future work to develop a more detailed understanding of the underlying mechanism that sets the magnitude of vegetation band curvature in the topographically-extended Klausmeier model as well as in more hydrologically accurate modeling frameworks may provide additional insight into how terrain shapes vegetation patterns via water transport.  Extension of the approach taken in~\cite{scheel2018advection} to two spatial dimensions may prove to be a useful path towards this goal.  The results may also be of interest to band formation in other contexts, particularly where experiments that test driving forces and explore ecological impacts have been possible~\cite{van2008experimental}. 

Comparison between the predictions based on the topographically-extended model and our observations from Australia, the United States and Ethiopia in Sec.~\ref{sec:discussion} shows qualitative agreement with respect to where patterns occur relative to topographic features such as ridges and valleys. 
 Limited simulations of a simplified version of the Gilad model (see Appendix~\ref{app:gilad}) indicate that these predictions obtained within the Klausmeier modeling framework (e.g. Fig.~\ref{fig:parspace}) remain qualitatively unchanged by subdividing the water field into surface flow and soil moisture. A more complete understanding of the extent to which results obtained with the highly conceptual Klausmeier model generalize to more hydrologically accurate models is required. 
Of particular interest 
are modeling frameworks that can resolve the fast timescales of rainfall events and surface waterflow dynamics, as well as relatively slower vegetation and soil moisture dynamics~\cite{konings2011drought,siteur2014will}. For instance, similar results regarding the influence of terrain curvature on pattern shape, location and migration speed have been reported by a more mechanistically detailed modeling study~\cite{baartman2018effect}. While this detailed model captures the intermittent nature of water input and the possible influence of vegetation patterns on the underlying terrain, its complexity makes it challenging to extract the dominant process by which the topography influences the vegetation patterns.

In this study we have focused on the role of topography in shaping vegetation bands, but many other mechanisms and environmental heterogeneities are likely to be involved as well. For example biomass feedback on water transport, boundary effects imposed at the edges of the patterns, spontaneous band break-up and interactions between different species of vegetation may all play an important role in the pattern formation process. Using studies in the context of peatland pattern formation as a template~\cite{eppinga2009nutrients,larsen2014exploratory}, we would like to construct and probe simple modeling frameworks that can capture other possible mechanisms in order to identify differences in predictions. This could suggest ways to leverage spatial heterogeneity in observational data to help determine dominant mechanisms at play. Exploring currently available data for aspects of the phenomenon that have the potential to exhibit observable differences in response to inhomogeneity can, in turn, inform such future model investigations.

By investigating the arcing direction of vegetation bands within a simple modeling framework and with an appropriately idealized  terrain, we have identified a possible link between band curvature and ecosystem fragility. 
Specifically, our model study suggests that convex upslope vegetation bands, confined to the lower ground of a landscape, may be closer to a collapse under increased aridity stress than convex downslope bands on the higher ground of a gently rolling terrain.   

\vspace{5mm}

\noindent\textbf{Data Accessibility.} All data is publicly available~\cite{DRUSCH201225,gallant2011adaptive,Farr:2007ib,ZOMER200867,gesch2002national,tadono2014precise}. 

\noindent \textbf{Author's Contributions.}  PG conceived the modeling approach, carried out the numerical analysis and took the lead in drafting the manuscript.  All authors participated in the data analysis and interpretation, contributed to the writing of the manuscript and gave final approval for publication. 

\noindent \textbf{Competing Interests.}  We have no competing interests.

\noindent \textbf{Funding.} This work was supported in part by the National Science Foundation grant DMS-1440386 to the Mathematical Biosciences Institute (PG), National Science Foundation grants DMS-1517416 (MS) and DMS-1547394 (KG), and the James S. McDonnell Foundation (KG).

\noindent \textbf{Acknowledgements.} We are grateful to Justin Finkel for helpful discussions. 


\appendix

\section{Detailed simulation results for topographically extended Klausmeier model}
\label{app:sim}

We use the idealized terrain given by Eq.~\eqref{eq:modelev}, fix the plant mortality to $m=0.45$~\cite{klausmeier1999regular}, and explore the influence of precipitation $a$ and channel aspect $\sigma$ on patterns within the topographically-extended Klausmeier model.  Starting with small ($\sim$ 5\%) spatially uncorrelated Gaussian noise on top of a uniformly vegetated initial condition, we integrate forward in time using a fourth order exponential time differencing scheme~\cite{cox2002exponential,kassam2005fourth} on an equidistributed mesh. Our calculations are performed in Fourier space and are fully dealiased. We use a periodic domain of  $[0,50]\times[0,200]$ on a  $128\times 512$ grid.  We consider our domain to be a small section of a hillslope and therefore employ periodic boundary conditions along the $x$-direction. While the elevation function $\zeta$ in Eq.~(\ref{eq:modelev}) is not actually periodic along the $x$-direction because it decreases steadily downslope, the use of periodic boundary conditions is nevertheless readily implemented since only derivatives of $\zeta$ appear in Eq.~(\ref{eq:transW}).

Generalizations of the Klausmeier model with transport $T^D_{vx}(W)=d\nabla^2(W^\gamma)+vW_x$ that include (potentially nonlinear) water diffusion in addition to the existing advection term in $T^K_{vx}(W)$ allow for pattern formation  even when $v=0$~\cite{van2013rise} and, as noted in Sec.~\ref{sec:intro}, lead to spontaneous break-up of stripes  on uniformly sloped terrain through secondary instabilities~\cite{siero2015striped}.  The Gilad model~\cite{gilad2004ecosystem} uses a water transport term of the form $T^G_\zeta (H)=\nabla \cdot (H\nabla (\zeta+H))$ for surface water $H$, which also includes a nonlinear water diffusion contribution that is not present for $T_\zeta(W)$. In this study, we focus on the influence of topography in the context of banded patterns where water transport is thought to be advection-dominated~\cite{valentin1999soil}.  We therefore do not expect water diffusion to play an essential role and, indeed, we find that including a (linear or nonlinear) diffusion term in $T_\zeta(W)$ has negligible effect on the results of our simulations provided some biomass is present.  Heuristically, one possible danger with not including some form of water diffusion in the model is that water may be able to accumulate in regions of high curvature faster than it can evaporate. However we are modeling water-limited ecosystems and capturing dynamics on time scales of years or longer.  Therefore the water input is assumed to be a constant mean value instead of having large spikes corresponding to rain events. In this situation we find that the biomass can effectively use up water and therefore diffuses fast enough to prevent any divergence.

\subsection{Arcing of bands in shallow channels}\label{sec:model:shallow}

\begin{figure*}
\includegraphics[width=\textwidth]{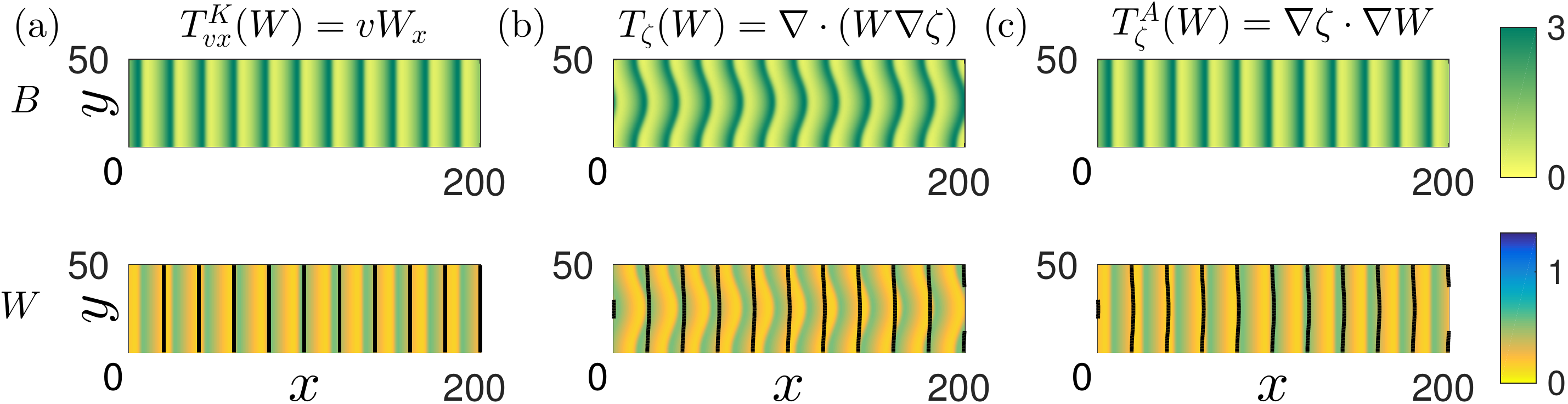}
\caption{  \label{fig:accumulation}
Biomass ($B$) and water ($W$) at $t=1000$ from a simulation of the Klausmeier system, Eqs.~\eqref{eq:klaus:W}-\eqref{eq:klaus:B}, (a) initialized with uniform vegetation and small noise using  $T^K_{vx} (W)= v W_x$ and (b)  $T_\zeta (W) =\nabla  \cdot (W \nabla\zeta )$ where $\zeta=v \left( x + \sigma \cos (k_0 y) \right)$.  The solution shown in (b) is taken as initial condition for a simulation using  $T^A_\zeta (W) =\nabla\zeta \cdot \nabla W$ and the result at $t=1000$ is shown in (c).  For each case black lines of constant elevation (straight in (a) and slightly curved in (b,c)) are superimposed on $W$.   If the $W\nabla^2\zeta$ term is neglected from $T_\zeta(W)$, then the simulation tends to  straight vegetation bands shown in (c). The simulations are carried out on a periodic domain of size $200\times 50$ with parameters $a=0.95$, $v=10$, $m=0.45$, $\sigma = 0.5$, $k_0=2\pi/50$. 
}
\end{figure*}

When a small channel aspect ($0<\sigma \lesssim 0.6$) is taken,  simulations produce traveling wave patterns with bands that arc in the same direction as the underlying elevation contours. The straight band traveling solutions shown for the Klausmeier water transport $T^K_{vx}(W)$ in Fig.~\ref{fig:accumulation}(a) become the modulated bands shown in Fig. 5(b) when the topographic extension $T_\zeta(W)$ with $\sigma=0.5$ is used.  While the bands are modulated such that the arcing-direction matches the direction of curvature of the underlying terrain along ridges and valleys, the vegetation bands do not closely match the contour lines shown superimposed on the water field $W$. There is only a slight curvature of the contour lines in Fig.~\ref{fig:accumulation}(b) while the bands are noticeably more arced.

We also consider an alternate extension to the Klausmeier transport $T^K_\zeta (W)$ that includes only the advection term of Eq.~\eqref{eq:transWadac}. We find that this advection-only transport,
\begin{equation} \label{eq:transWnaive}
T^A_\zeta(W)=\nabla \zeta \cdot \nabla W,
\quad \zeta=v(x+\sigma \cos(k_0y)),\end{equation} 
does not produce arced vegetation bands. Figure~\ref{fig:accumulation}(c) shows the result of a simulation initialized with arced vegetation bands from Fig.~\ref{fig:accumulation}(b) and using the
transport $T_\zeta^A(W)$ above.  The initially curved vegetation bands straighten out to become the $y$-independent solution of the original Klausemeier model shown in Fig.~\ref{fig:accumulation}(a). 
The persistence of this straight-band solution in the presence of the advection-only transport  $T_\zeta^A(W)$ for $\sigma\neq 0$ can be understood as follows. A $y$-independent solution to the original Klausmeier system with transport $T_{vx}^K(W)$ will also be a solution to the system with advection-only transport $T_\zeta^A(W)$.  This is because $\nabla W=W_x \hat{\mathbf{x}}$ if $W_y=0$, and  so $T_\zeta^A(W)=T_{vx}^K(W)$ for $y$-independent $W$. Thus the solution with straight bands aligned along $y$ continues to exist, and numerical simulations (Fig.~\ref{fig:accumulation}(c)) indicate that this straight-band solution not only remains stable for $\sigma \neq 0$, but is actually a preferred solution in this example.

\subsection{Classifying patterns on ridges and in valleys}\label{sec:model:classify}

\begin{figure*}
\includegraphics[width=\textwidth]{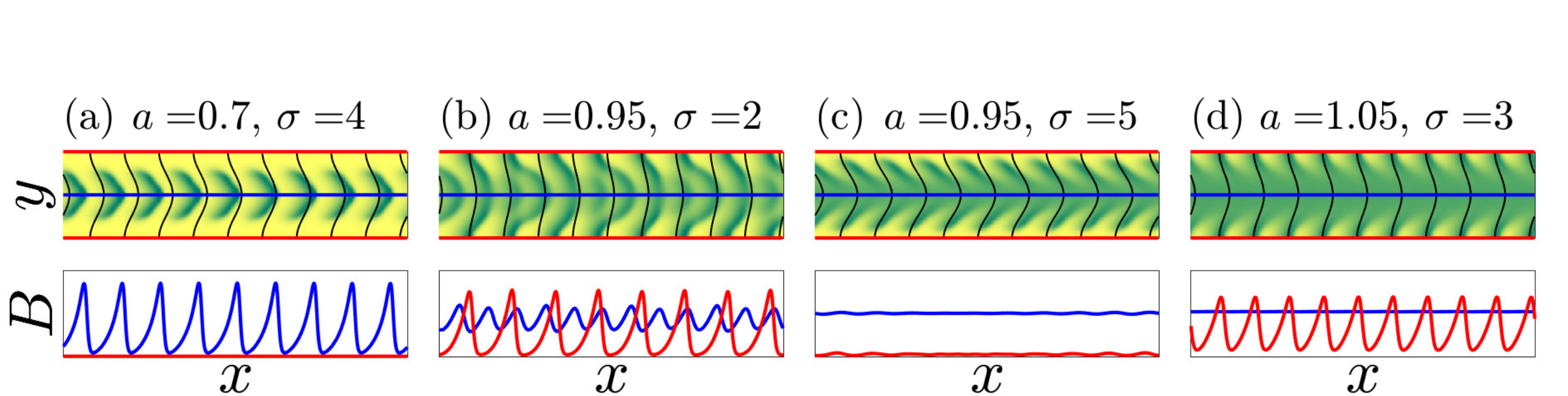}
\caption{  \label{fig:preview}  Qualitatively different patterns  as a function of the parameters $(a,\sigma)$. Biomass is shown at t=1000 from simulations with (a) $a=0.70$ and $\sigma=4$, (b) $a=0.95$ and $\sigma=2$, (c) $a=0.95$ and $\sigma=5$, and (d) $a=1.05$ and $\sigma=3$. For each case, the upper image shows the biomass on the full 2D domain while the lower image shows profiles of the biomass along the ridge (red) and valley (blue). 
Simulations are initialized with small noise on top of a uniformly vegetated state and carried out on a periodic domain of size $200\times 50$ with parameters $v=10$, $m=0.45$, $k_0=2\pi/50$.  
}
\end{figure*}

Figure~\ref{fig:preview} shows snapshots of biomass (at $t=1000$) from simulations along with elevation contours for various values of precipitation $a$ and channel aspect ratio $\sigma$ that illustrate qualitatively different patterns predicted by the model.
We choose a precipitation value ($a=0.95$) for which stable traveling waves of banded vegetation moving uphill exist on a uniformly sloped terrain ($\sigma=0$ as in Fig.~\ref{fig:accumulation}(a)) and introduce a curvature in elevation transverse to the slope ($\sigma>0$).  As noted in Sec.~\ref{sec:model:shallow}, e.g., $\sigma= 0.5$ in Fig.~\ref{fig:accumulation}(b), the asymptotic state remains a traveling wave with velocity directed uphill for small aspect ratio $\sigma$, but the bands are arced convex-upslope within valleys and convex-downslope on top of ridges.  

For $\sigma=2$, shown in Fig.~\ref{fig:preview}(b), the simulation produces a periodic pattern with a wavelength that is 1.5 times longer on the ridges than in the valleys.  The speed of the pattern on ridges is slower than within valleys, resulting in phase slips in which bands extending across the ridges and valleys break apart into arced segments and reconnect with other arced segments to form new bands that extend across the domain.  We note that the speed can be either faster or slower in a valley, depending on parameters and the ratio of wavelengths.  When the wavelengths are the same, as is the case when $\sigma=1$ and $a=0.95$ (not shown), we find that the pattern in the valleys moves faster than the pattern on the ridges.  When the wavelengths are different, we find cases (e.g., $\sigma=2$ and $a=0.95$ shown in Fig.~\ref{fig:preview}(b)) where the longer wavelength pattern on ridges moves faster than the shorter wavelength pattern in valleys. Such behavior is not unexpected based on trends of linear theory: the speed of uphill migration computed from linearization about the uniform state increases with increasing wavelength and decreases with increasing water loss rate when all other parameters are fixed~\cite{sherratt2005analysis}.  
Eventually, for large enough aspect $\sigma$, nearly uniform vegetation cover develops within  valleys while ridges becomes nearly bare (for example $\sigma=5$ in Fig.~\ref{fig:preview}(c)). In this case the vegetation bands are restricted to the channel walls between the valley and ridge lines, forming chevrons.

\subsection{Mapping out patterns in the $(a,\sigma)$-plane}

We map out the existence of patterned states of the topographically-extended Klausmeier model in the $(a,\sigma)$-plane by scanning over the precipitation parameter $a$ for fixed values of channel aspect $\sigma$ as described below. We initialize each simulation with a uniformly vegetated state at a high enough precipitation value $a$ so as to ensure that a state consisting of uniform vegetation cover on both ridges and valleys is stable at the given value of $\sigma$.  For each fixed channel aspect $\sigma$, we decrease the value of $a$ at a rate $da/dt=-5\times 10^{-5}$, adding 1\% spatially uncorrelated Gaussian noise to the solution at time intervals of $\Delta t=100$.  As a check for hysteresis, at each fixed value of $\sigma$ we also run the simulation for increasing precipitation ($da/dt=+5\times 10^{-5}$) starting from the last value of $a$ before the vegetation in the system collapses to the bare soil state.

We consider biomass profiles along the ridge $B_R(x)=B(x,y=0$) and the valley ($B_V(x)=B(x,y=L/2$) (red and blue profiles in Fig.~\ref{fig:preview}) for a grid in the $(a,\sigma)$ plane with spacing $\Delta a=0.01$, $\Delta \sigma =0.5$. At each point on the parameter grid, the pattern amplitude along the ridge ($R$) and valley ($V$) lines is computed via $A_{R,V}=\mathrm{max}_x(B_{R,V}(x))-\mathrm{min}_x(B_{R,V}(x))$. Because of hysteresis, the amplitudes $A_{R,V}$ may be different for the decreasing and increasing $a$ portions of the simulation at a given point in parameter space. We are interested in mapping out the parameter region of existence for patterns so we take the maximum of $A_{R,V}$ between the increasing and decreasing $a$ cases, whenever a difference can be computed.

\begin{figure*}
\includegraphics[width=\textwidth]{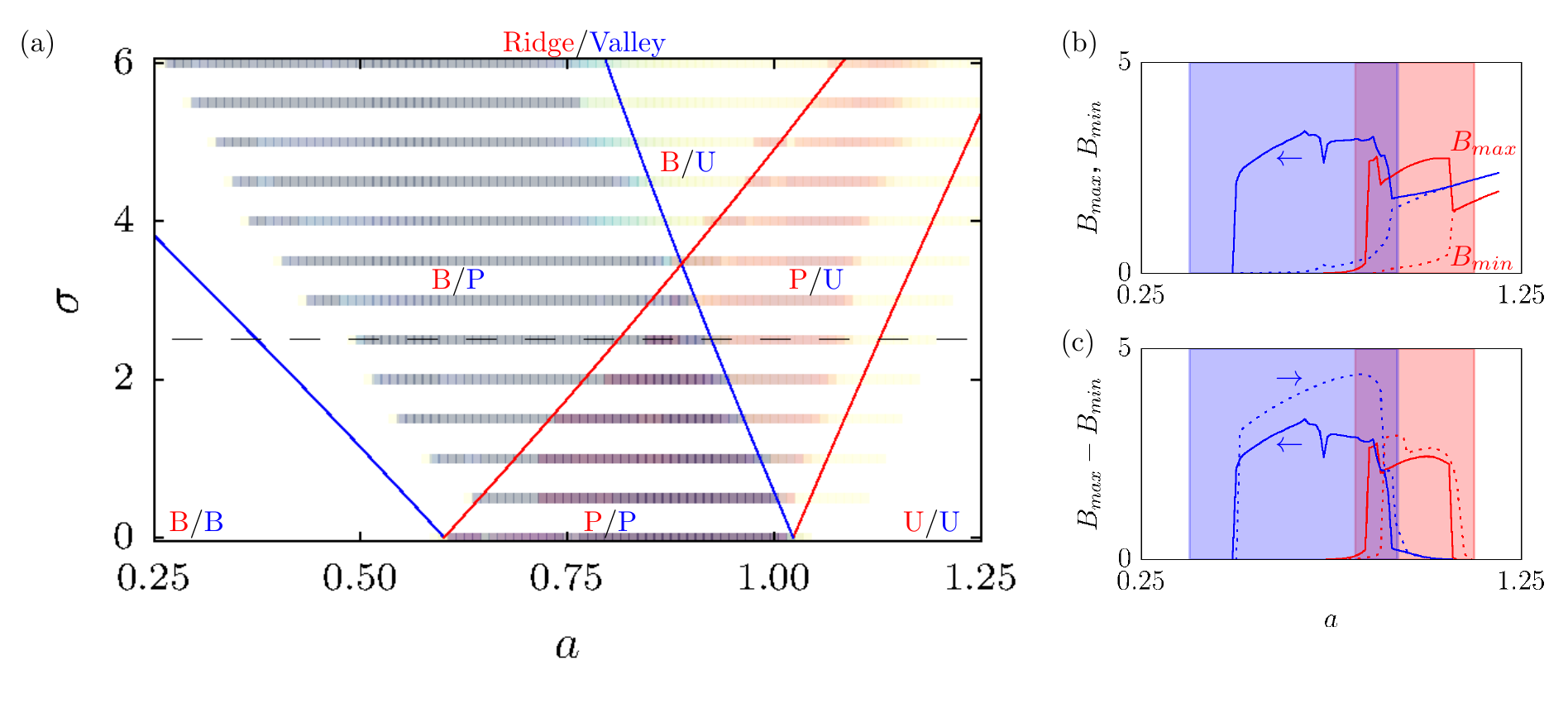}
\caption{\label{fig:appparspace} (a) Maximum of amplitude of patterns on ridge (red) and valley (blue) in $(a,\sigma)$ plane when $a$ is decreased and then increased for fixed $\sigma$.  Yellow indicates small amplitude while red (blue) indicates large amplitude on ridge (valley) and purple indicates large amplitude on ridge and in valley.  Solid red (blue) lines show predicted transitions between regions of parameter from the one-dimensional approximation for the ridges (valleys) given by Eq.~(\ref{eq:klaus1d:W})-(\ref{eq:klaus1d:B}).  Each region is labeled by ridge/valley state as bare soil (B), patterned (P) or uniformly vegetated (U).  (b) Profiles of $\mathrm{max}_x(B)$ (solid) and $\mathrm{min}_x(B)$ (dotted) along the ridge (red) and valley (blue) are shown for decreasing   $a$ for fixed $\sigma=2.5$ (dashed line in (a)). (c) Profiles of pattern amplitude $A=\mathrm{max}_x(B)-\mathrm{min}_x(B)$ along ridge (red) and valley (blue) for $a$ decreasing (solid) and increasing (dotted) at fixed $\sigma=2.5$.  The shaded regions in (b) and (c) indicate values of $a$ where the one-dimensional approximation predicts patterns to exist on the ridge (red) and in the valley (blue).  Parameters: $m=0.45$, $v=10$, $k_0=2\pi/50$. }
\end{figure*}

 Figure~\ref{fig:parspace}(a) shows the maximum ridge (valley) pattern amplitudes $A_R$ ($A_V$) in red (blue) at each point in the $(a,\sigma)$-plane.   Purple indicates where patterns appear on both ridges and valleys while yellow  indicates where no significant pattern amplitudes appear on either ridges or valleys.  The visualization of the data in Fig.~\ref{fig:parspace}(a)  does not distinguish between uniform vegetation cover and bare soil so we additionally label each region of the parameter space based on the ridge and valley states.  The algorithm for obtaining the amplitude is illustrated for $\sigma=2.5$ (see dashed gray line) in the right panels.
Figure~\ref{fig:appparspace}(b) shows $\mathrm{max}_x(B(x))$ (solid) and $\mathrm{min}_x(B(x))$ (dotted) as the parameter $a$ is decreased for $y=0$ (red), corresponding a ridge,  and  for $y=L_y/2$ (blue), corresponding to a valley.  The pattern amplitude, computed by taking the difference, $A=\mathrm{max}_x(B(x))-\mathrm{min}_x(B(x))$, are shown for ridge (red) and valley (blue) in Fig.~\ref{fig:appparspace}(c). Here the solid line represents the pattern amplitude as $a$ is decreased and the dotted line represents the pattern amplitude as $a$ is increased.  The shaded regions in panels (b) and (c) represent the predictions for the existence of patterns from the one-dimensional model for the ridge (red) and the valley (blue).

\section{Simulation results from a three-field model}\label{app:gilad}
We have partially mapped out the influence of terrain curvature as a function of water input using a three-field model that subdivides water  into a surface water field $H$ and a soil moisture field $W$.  The model we use, a simplified version of the Gilad model~\cite{gilad2004ecosystem}, takes the form:  
\begin{align}
H_t &= p-I + D_h\nabla^2 H + \nabla \zeta \cdot \nabla H+ (\nabla^2\zeta) H\\
S_t &= I - \nu S (1-\rho B) - \gamma G_w\\
B_t &= -m B + \nu G_w (1-B)+D_b\nabla^2 B
\end{align}
where infiltration and transpiration functions are given by
\begin{equation}
I=\alpha H \frac{B+q f}{B+q},\qquad G_w= W B (1+\eta B)^2
\end{equation}
and the elevation function is
\begin{equation}
\zeta=v \left( x+\sigma \cos(k_0 y) \right)
\end{equation}
The parameter values used for the simulations are: $D_b=1$, $D_h=10$, $v=50$, $\nu=3.33$, $\rho=0.95$, $m=1.0$, $\gamma=16.66$, $\alpha=33.3$, $q=0.05$, $f=0.1$,  $\eta=3.5$.

\begin{figure*}
	\includegraphics[width=\textwidth]{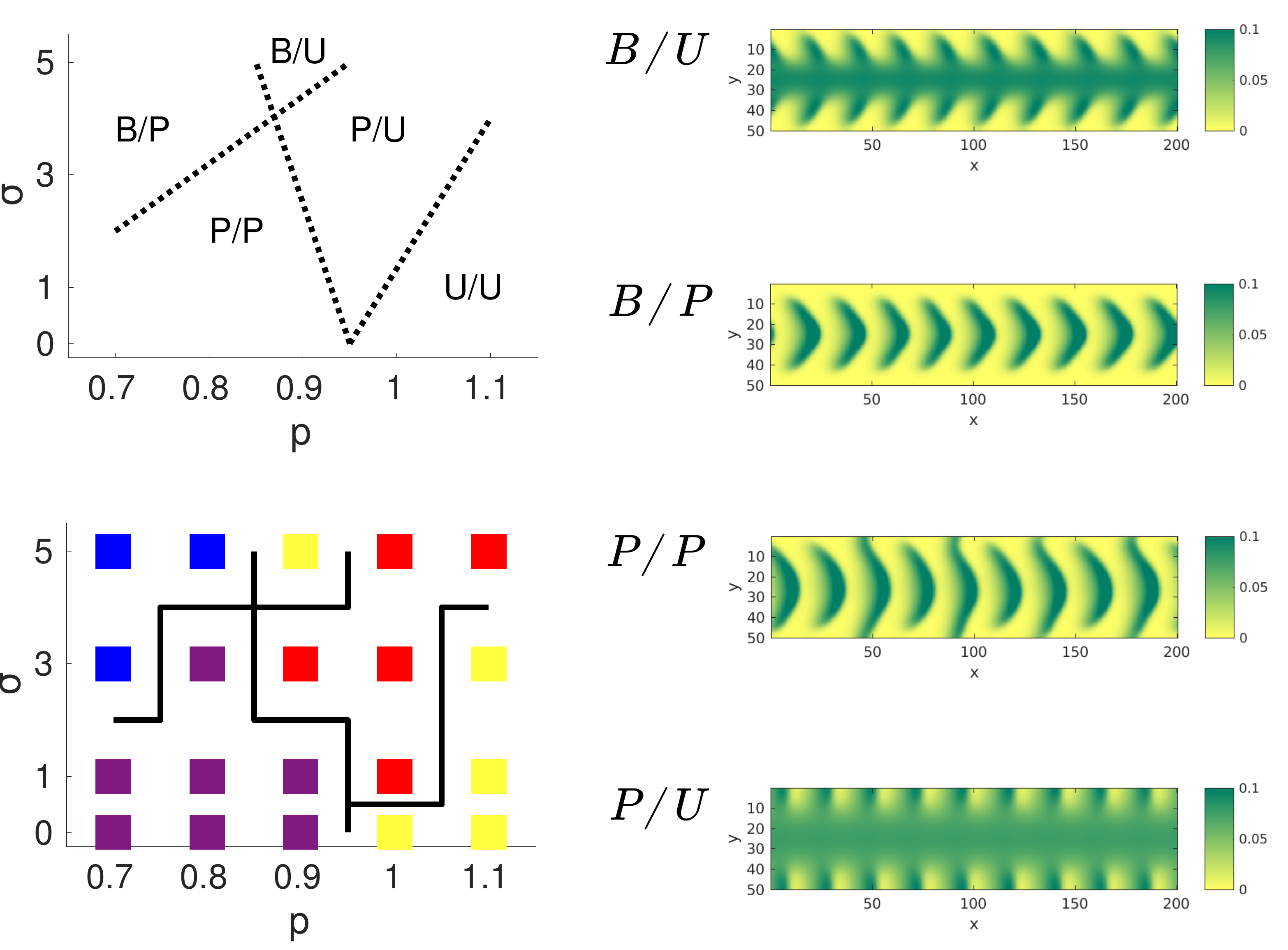}
	\caption{\label{fig:gilad} Simulation results based on a local approximation to the Gilad model using an idealized terrain consisting of a periodic array of ridges and valleys aligned along a hillslope. The precipitation level is characterized by $p$ and the terrain curvature is parametrized by $\sigma$.  Upper left panel: A cartoon indicating where in the $( \sigma , p)$-plane ridge/valley states are bare soil (B), vegetation patterns (P), and uniform vegetation cover (U).  Lower left panel: Squares are colored to indicate the ridge/valley state observed from numerical simulation for a given precipitation $p$ and terrain curvature  $\sigma$. Blue: B/P, Red: P/U, Purple: P/P, and Yellow: B/U or U/U.  Right Panels: Example biomass profiles for pattern types appearing in the simulation.  
		}
\end{figure*}

Figure~\ref{fig:gilad} summarizes the results of these simulations carried out on a two-dimensional domain in the $(p,\sigma)$-plane.      
The patterns generated from the simulation, once classified in terms of ridge/valley vegetation state, reveals a qualitatively similar parameter space structure as seen for the topographically extended Klausmeier model in Fig.~\ref{fig:appparspace}.  While we do not present numerical continuation results of a one-dimensional Gilad model, the structure of the equations allows us to interpret terrain curvature as effectively changing water loss rate in the equation for surface water $H$ in a similar way as was done for the equation for water $W$ in the Klausmeier model.   Because the surface water is diverted from ridges and accumulates in valleys, we see an increased infiltration  and thus increased level of soil moisture in valleys relative to ridges.

\bibliographystyle{unsrt}
\bibliography{arc}
\end{document}